\documentclass[fleq,usenatbib]{mnras}
\usepackage{newtxtext,newtxmath}
\usepackage[T1]{fontenc}
\usepackage{ae,aecompl}
\usepackage{color,soul}
\usepackage{graphicx}
\usepackage{amsmath}
\usepackage{amssymb}
\usepackage{color}
\newcommand{\acro}{TREVR}
\providecommand{\e}[1]{\ensuremath{\times10^{#1}}}
\newcommand{\bigO}[1]{\mathcal{O}\left(#1\right)}

\newcommand{\NS}{N_{\rm source}}
\newcommand{\NR}{N_{\rm ray}}
\newcommand{\NK}{N_{\rm sink}}
\newcommand{\tr}{\tau_{\rm ref}}
\newcommand{\tO}{\theta_{\rm open}}
\newcommand{\strom}{Str\"omgren}

\title[]{\acro{}: A general $N\log^2N$ radiative transfer algorithm}

\author[J. J. Grond et al.]{
J. J. Grond,
R. M. Woods,
J. W. Wadsley \thanks{E-mail:  wadsley@mcmaster.ca}
and H. M. P. Couchman
\\
Department of Physics and Astronomy, McMaster University, Hamilton, Ontario L8S
 4M1, Canada}

\date{Accepted 2019 February 15. Received 2019 January 25; in original form 2018 October 10}

\pubyear{2019}

\begin{document}
\label{firstpage}
\pagerange{\pageref{firstpage}--\pageref{lastpage}}
\maketitle

\begin{abstract}
We present \acro{} (Tree-based REVerse Ray Tracing), a general algorithm for 
computing the radiation field, including absorption, in astrophysical 
simulations. \acro{} is designed to handle large numbers of sources and 
absorbers; it is based on a tree data structure and is thus suited to 
codes that use trees for their gravity or hydrodynamics solvers (e.g. Adaptive 
Mesh Refinement). It achieves computational speed while maintaining a 
specified accuracy via controlled lowering of the resolution of both sources 
and rays from each source. \acro{} computes the radiation field in order
${N\log\NS}$ time without absorption and order ${N \log \NS \log{N}}$ time 
with absorption. These scalings arise from merging sources of radiation 
according to an opening angle criterion and walking the tree structure to 
trace a ray to a depth that gives the chosen accuracy for absorption. The 
absorption-depth refinement criterion is unique to \acro{}. We provide a suite 
of tests demonstrating the algorithm's ability to accurately compute fluxes, 
ionization fronts and shadows.   

\end{abstract}




\section{Introduction}\label{sec:intr}
Radiation, arguably, plays the determining role in the field of astrophysics. 
Almost all of the information we receive from the cosmos comes in the form of 
photons we detect on or around earth. Understanding the process of radiative 
transfer (RT) is required to interpret this information, as the photons are 
affected by the media they travel through on their way to our telescopes and 
detectors. Interactions between photons and these media not only affect the 
photons themselves but the matter as well. Photons and baryons exchange energy 
and momentum, driving both heating and cooling. This also affects excitation 
and ionization states and thus determines the chemical and thermodynamic 
properties of the gas. Thus radiation is a key player in many of the 
astrophysical systems and processes we study.

\begin{figure}
\includegraphics[width=1\linewidth]{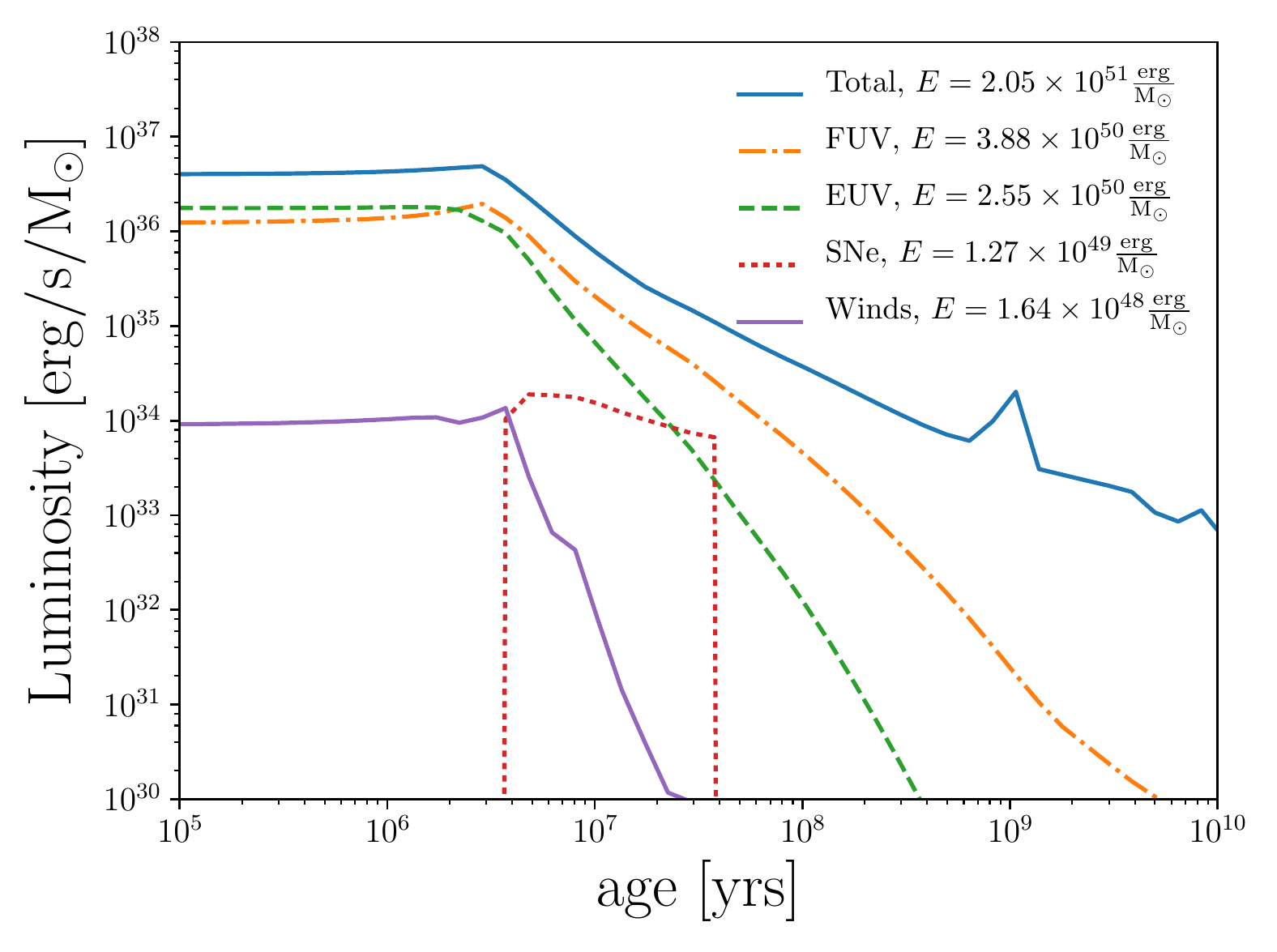}
\caption{Luminosity per solar mass as a function of time for a stellar 
population having a Chabrier initial mass function \citep{chabrier03}.}
\label{fig:uvsn}
\end{figure}

On galaxy scales, a central question is how feedback mechanisms affect star 
and galaxy formation. Stellar feedback comes in the form of photoionization by 
ultraviolet (UV) radiation, stellar winds and supernovae 
\citep[e.g.][]{leithererEt99}, the latter of which has been a main focus in 
simulations in previous years \citep[e.g.][]{agertzEt13}. It is important to 
note that even though supernovae might be spectacularly powerful events, 
ionizing radiative output from stellar populations contributes two orders of 
magnitude more energy at early times and about 50 times more energy over the 
course of a stellar population's lifetime. This is evident from 
Figure~\ref{fig:uvsn}, which is a plot of the luminous output per solar mass 
as a function of time from a typical stellar population (computed using the 
stellar evolution code Starburst99; \citealt{leithererEt99}).

However, the way in which this massive output of UV radiation is deposited 
and consequently affects the interstellar medium (ISM) is still unclear. 
Attempts at numerically exploring these effects without the use of a full 
radiative transfer method have produced conflicting results. Simulations done
by \cite{gritschnederEt09} and \cite{walchEt12} suggest that ionizing feed 
back from large O-type stars before the first supernovae ($\sim 1-3\ \rm{Myr}$) 
have a significant effect on star formation rate. Whereas \cite{daleEt12} 
conclude the effect on the star formation rate to be small.

With this potential impact in mind, it may seem surprising that RT has been 
treated poorly in most galaxy-scale astrophysical simulations, often as an 
imposed uniform background. This is because RT is an intrinsically complex and 
computationally expensive problem. The complexity is immediately evident from 
the full RT equation \citep[e.g.][]{mihalasMihalas84},
\begin{eqnarray} \label{eqn:classicrt}
\left[ \frac{1}{c} \frac{\partial}{\partial t} + \mathbf{n \cdot \nabla}
 \right] I\left(\mathbf{x}, \mathbf{n}, t, \nu\right) = \nonumber \\
\epsilon\left(\mathbf{x}, \mathbf{n}, t, \nu\right) - 
\alpha\left(\mathbf{x}, \mathbf{n}, t, \nu\right) 
I\left(\mathbf{x}, \mathbf{n}, t, \nu\right).
\end{eqnarray} 
Here, $I$, $\epsilon$ and $\alpha$ are the intensity, emissivity and 
extinction coefficients respectively and all depend on position $\mathbf{x}$, 
unit direction of light propagation $\mathbf{n}$, time $t$ and frequency 
$\nu$. Apart from being a seven dimensional problem, RT involves the highest 
possible characteristic speed, $c$, the speed of light. Also, unlike a similar
problem such as gravity, RT depends on the properties of the 
intervening material via the absorption term, $\alpha$.

Because of this complexity, a na\"ive numerical solution to the RT problem 
scales with the number of resolution elements, $N$, as $\bigO{N^{7/3}}$ and 
requires a timestep thousands of times smaller than typical Courant times in 
astrophysics. This scaling arises due to three contributions. Firstly, a 
radiation field must be computed at each of the simulation's $N$ resolution 
elements. Secondly, each one of the resolution element's intensity values is 
made up of contributions from $\NS$ sources of radiation ($\NS$ rays of light 
being computed per resolution element). This leads to a scaling for the total 
number of rays of $\NR = N \times \NS$, or $\bigO{N^2}$ assuming that $\NS 
\sim N$. This fact alone limits brute-force RT methods to only small-scale 
problems, such as ionization by a few massive stars \citep[e.g.][]{howard16, 
howard17}. Finally, each ray of light interacts with the medium along its 
path, which is resolved with $\bigO{N^{1/3}}$ resolution elements. Thus the 
computational cost is $\bigO{N^{7/3}}$. This poor scaling with number of 
resolution elements makes it infeasible, or at least unattractive, to simulate 
RT alongside gravity and hydrodynamics methods that scale as $\bigO{N\log N}$ 
or better. It is evident that much can be gained by reducing the linear 
dependence on $\NS$, with additional gains from tackling the $N^{1/3}$ cost 
per ray.

A practicable RT method would have to solve a simplified RT problem. RT 
methods can be divided into two different categories based on how they treat 
$c$ in Equation~\ref{eqn:classicrt}.

\textit{Evolutionary} methods use a 
finite $c$, (which is often reduced from the true speed of light) and thus the 
partial time derivative remains in Equation~\ref{eqn:classicrt}, and the 
radiation field is advected or evolved thoughout the simulation. The 
prototypical evolutionary method is flux-limited diffusion (FLD) 
\citep{levermorePomraning81}. Modern evolutionary methods include moment 
methods like OTVET \citep{gnedinAbel01} and RAMSES-RT \citep{rosdahlEt13} as 
well as photon packet propagation methods like TRAPHIC \citep{pawlikSchaye08}, 
SPHRAY \citep{altayEt08} and SimpleX2 \citep{paardekooperEt10}.

\textit{Instantaneous} methods, on the other hand, take the limit where $c$ 
is infinite and the partial time derivative in Equation~\ref{eqn:classicrt} 
goes to zero. In this case the radiation field can be computed instantaneously 
as a geometric problem. Computational methods in this category include forward 
raytracers such as $\rm C^2Ray$ \citep{mellemaEt06a}, Moray 
\citep{wiseAbel11} and Fervent \citep{baczynskiEt15} as well as reverse 
raytracers such as TreeCol \citep{clarkEt12}, URCHIN \citep{altayTheuns13} 
and TREERAY \citep{Wunsch2018, HaidEt18}. 

Instantaneous methods typically take the form of raytracers. Raytracers are 
the most direct way to solve the RT problem. Forward raytracers trace many 
rays outward from sources of radiation, similarly to the actual phenomenon, in 
the hope that resolution elements will have sufficiently many rays 
intersecting them to compute a radiation field. Na\"ively, the number of rays 
per source would be comparable to the number of resolution elements, giving a 
scaling of $\bigO{N \NS N^{1/3}}$, as previously noted. However, for forward 
ray tracing, $\bigO{N^{2/3}}$ rays per source are typically sufficient to hit 
every resolution element when extended to the edge of the simulation volume 
(distance $\bigO{N^{1/3}}$), so the scaling typically achieved is 
$\bigO{N \NS}$.  

It is important to note when methods adaptively split rays (e.g. using 
Healpix \citealt{gorskiEt05} as in Moray, URCHIN and TreeCol), it does not 
change the overall scaling. For example, a centrally located source requires 
$6\, N^{2/3}$ rays to strike all elements in the outer faces of a cubical 
simulation volume, each with a length $\bigO{N^{1/3}}$. Even with adaptive ray 
merging near the source, at least $N$ ray segments are required to intersect 
each of the $N$ resolution elements. In addition, raytracers such as Moray 
rely upon a Monte-Carlo approach to estimate the radiation field and thus 
require at least 10 rays to intersect each element, a constant but 
significant prefactor to the overall cost. This scaling usually limits 
forward raytracers to problems with few sources to avoid 
$\mathcal{O}(N^2)$-like scaling. 

Recently there has been some focus on reverse ray tracing methods by
\cite{clarkEt12}, \cite{altayTheuns13}, \cite{Woods2015} (applied in 
\citealt{KannanEt14}) and \cite{HaidEt18}. The first two methods listed are 
not general, as they are designed to compute external radiation (e.g. from the 
post-ionization UV background) rather than internal sources of radiation.
The latter two methods are more general and can handle internal sources.

The idea of reverse ray tracing introduces some advantages relative 
to forward ray tracing. Reverse raytracers trace all the rays that strike
a specific resolution element before moving to the next element. 
Algorithmically, this is equivalent to tracing in reverse, from the sinks to 
the sources. This makes it easy to ensure that the source and absorber angular 
distributions are well-sampled near the resolution element as opposed to 
forward ray tracing where one would have to increase the number of rays per 
source to guarantee this type of accuracy. Put simply, radiation is computed 
exactly where it is needed.  This is especially advantageous in adaptive mesh 
and Lagrangian simulations such as smoothed particle hydrodynamics (SPH), as 
low density regions are represented by few resolution elements, and thus extra 
work is not done to resolve radiation in those regions.

A key benefit to reverse ray tracing is the potential for adaptive timesteps 
to dramatically reduce the radiation work as only active resolution elements, 
$\NK$, need to be traced to. This active subset can be a million times smaller 
than $N$ in, for example, high-resolution cosmological simulations. Typical 
hydro and gravity codes achieve a factor of 100 speed-up by taking advantage 
of this so it is important that the radiation code has the same capability or 
radiation will overwhelm the computation. Thus a na\"ive reverse ray trace 
still scales as $\bigO{\NK \NS N^{1/3}}$, with the presence of many sources 
presenting the most significant computational barrier.

In contrast, evolutionary methods are typically based on evolving moments of 
the radiation field stored at each resolution element. They are insensitive to 
the number of sources, and scale as $\mathcal{O}(N)$ with the number of 
resolution elements, allowing them to handle large numbers of sources and 
scattering. Although evolutionary methods can handle both optically thin and 
thick regimes, they lose directional accuracy in intermediate regimes and 
suffer from poor directional accuracy in general. This is immediately apparent 
in shadowing tests ({e.g. Figure~16 in \citealt{rosdahlEt13}).

Photon packet propagation methods, such as TRAPHIC \citep{pawlikSchaye08}, 
employ an evolutionary approach in which directional accuracy is easier to 
control, in principle. However, the Monte-Carlo aspects of how photon packets 
are propagated introduce significant Poisson noise into their computed 
radiation field. Added Monte-Carlo re-sampling is shown to reduce this noise 
but it is quite expensive and degrades the initially sharp shadows. For this 
reason  it is typically not used in production runs. TRAPHIC also adds virtual 
particles (ViPs) to propagate their photon packets in less dense, optically 
thin regions lacking in SPH particles. TRAPHIC scales linearly with resolution 
elements, as mentioned before, multiplied by the number of packets per element 
(typically 32-64).

A key limitation for evolutionary methods, whether they are moment or 
packet-tracing methods, is that the radiation field for every element needs to 
be computed every timestep. In addition, the speed of light, even when 
reduced, is substantially larger than the sound speed and thus many radiation 
substeps are required compared to the hydro solver. Thus for photon packet 
propagation methods every photon packet typically hops forward several times 
for each hydro step even if most elements are not active. A key outcome is 
that moment methods cannot take advantage of adaptive timesteps to limit 
radiation work.  Another issue, specific to TRAPHIC, is that $N$ is 
significantly greater than the number of SPH particles due to the addition of 
ViPs. These factors dramatically increase the prefactor on the scaling.  
Nonetheless, methods such as TRAPHIC represent an effective approach for large 
simulations that can handle a variety of regimes of optical depth.

Until now the poor scaling with source number, as $\bigO{\NS N}$, has severely 
limited the applicability and competitiveness of instantaneous ray tracing 
relative to evolutionary methods such as TRAPHIC. Recently however, 
\cite{Woods2015} and \cite{Wunsch2018} developed promising generalizations of 
reverse ray tracing based on merging of sources that can handle large numbers 
of internal sources. The basic idea is to use a tree to combine distant 
sources and reduce the cost to $\bigO{N \log N}$. \cite{Wunsch2018} implemented 
the TREERAY reverse raytracer in the FLASH AMR code \citep{FryxellEt2000}.
They employ an Oct-tree, a fixed number of rays (48) per source and calculate 
absorption on the fly during the tree-walk.  The primary weakness of the 
\cite{Woods2015} and \cite{Wunsch2018} methods is that they lower the resolution 
along rays in a preset manner.  This prevents them from maintaining the 
accuracy of the received flux at higher optical depths. Doing so requires multiple 
adaptivity criteria.  This means going beyond the open angle used in tree codes.
This is the focus of the current work.

In this paper we present \acro{} (Tree-based Reverse Ray Tracing), an $\bigO{N 
\log^2 N}$ adaptive reverse raytracer. In Section~\ref{sec:mthd} we detail 
the specific RT equations \acro{} solves (Subsection~\ref{sec:rteq}) and the 
general \acro{} algorithm (Subsection~\ref{sec:algo}) including its adaptivity 
criteria. \acro{} is not specific to any one kind of code (e.g. Adaptive Mesh 
Refinement (AMR) vs. SPH). Here we provide details of our implementation in 
the \textsc{Gasoline} SPH code (Subsection~\ref{sec:specs}). \acro{} was 
developed from the method of \cite{Woods2015} which was also implemented in 
\textsc{Gasoline}. In Section~\ref{sec:tsts} we present a suite of tests 
demonstrating the algorithm's ability to accurately compute fluxes, ionization 
fronts and shadows in the optically thick and thin regimes. These tests also 
allow us to explore how \acro{}'s adaptivity criteria control error and 
affects computational cost. The computational cost is bounded and 
characterized in the general case to substantiate the $\bigO{N \log^2 N}$ 
claim made earlier. Finally, in Section~\ref{sec:disc} we discuss \acro{}'s 
strengths and shortcomings and conclude how they enable and constrain the 
types of problems \acro{} can handle, and discuss improvements that can be 
made in the future.

\section{Method}\label{sec:mthd}
\subsection{Simplifications to the full RT problem}\label{sec:rteq}
Before describing \acro{}, we will first define the simplified version of the 
classical RT equation that the method solves. Since \acro{} is an instantaneous 
method, $c$ is set to infinity eliminating the partial time derivative in 
Equation~\ref{eqn:classicrt} leaving us with the instantaneous RT equation:
\begin{equation} \label{infcrt}
\mathbf{n \cdot \nabla} I\left(\mathbf{x}, \mathbf{n}, t, 
\nu\right) = \epsilon\left(\mathbf{x}, \mathbf{n}, t, \nu\right) - 
\alpha\left(\mathbf{x}, \mathbf{n}, t, \nu\right) 
I\left(\mathbf{x}, \mathbf{n}, t, \nu\right).
\end{equation}
The emissivity term in the above equation, $\epsilon$, describes a continuous 
emitting medium. \acro{} could assume sources of radiation were continuous, but 
being a numerical method it needs to represent sources of radiation as 
discrete resolution elements such as ``star particles''. In this case $\epsilon$ is a sum of delta 
functions and the solution to the RT equation becomes a linear combination of 
contributions from all sources of radiation. Also, since we are considering 
sources one by one we can start using the path length $s$ between a source and 
resolution element as our integration element and examine just one direction, $\mathbf{n}$,
\begin{equation}
\label{eqn:combtransfer}
\frac{dI}{ds} = -\alpha I.
\end{equation}
We can then combine the path length and extinction coefficient to solve for 
intensity by integrating 
\begin{equation}
\label{eqn:dtau}
d\tau = \alpha ds = \kappa \rho ds, 
\end{equation}
for $\tau$, the optical depth, where $\kappa$ is opacity and $\rho$ is 
density. This leads to
\begin{equation}
\label{eqn:absorbtransfer}
\frac{dI}{d\tau} = -I,
\end{equation}
which is the final version of the RT problem solved by this method. The 
solution to the equation is 
\begin{equation}
\label{eqn:ient}
I(s) = I(0)e^{-\tau(s)},
\end{equation}
where $I(0)$ is the intensity of the source and $\tau(s)$ is the
quantity to be estimated in our method:
\begin{equation}
\label{eqn:tauint}
\tau(s) = \int_{0}^s \kappa(s) \rho(s) ds.
\end{equation}

If we assume that the source of radiation is point-like, then the intensity at 
the receiver (the sink) is a delta function in angle. In this case, there is a 
one-to-one correspondence between the intensity and flux contributions due to 
that source. The flux is given by 
\begin{equation}
  \mathbf{F} = \int I(\Omega) \mathbf{n'}(\Omega) d\Omega = I(s) \mathbf{n},
\end{equation}
where $\mathbf{n}$ is the unit vector in the direction from the source to the 
sink.

For each source, $i$, we have a luminosity, $L_i$, which can be directly 
converted to a contribution to the flux at the sink,
\begin{equation}
\label{eqn:simpflux}
\mathbf{F_i} = \frac{L_i}{4\pi s_i^2} e^{-\tau_i} \mathbf{n_i},
\end{equation}
where $\tau_i$ is the accumulated optical depth along the ray between that 
source and the sink and $s_i$ is the distance. The net flux, $\mathbf{F}$, is 
then computed by summing up flux contributions from all sources.

The intensity due to a single source is,
\begin{equation}
\label{eqn:simpint}
I_i = \frac{L_i}{4\pi s_i^2} e^{-\tau_i}.
\end{equation}
By summing the intensity from all sources we can get the angle-averaged 
intensity. We can use this averaged intensity directly in heating, chemistry 
and ionization rate expressions. For many applications in astrophysics this is 
the primary effect of the radiation field on local gas. We note that it is important
to apply timestep limits to correctly follow the progress of ionization fronts as they change
the absorption properties of the gas.  This approach
relies on the ionization front speed as the rate limiting step, rather than the speed of light.

The first order moment of the intensity is the net radiation flux.  Higher 
order moments such as the radiation pressure, $\mathbf{p}$, can be easily 
obtained with simple summations.

\subsection{Algorithm}\label{sec:algo}
The \acro{} algorithm is based around a tree data structure which partitions 
the simulation volume hierarchically in space. The smallest resolution 
elements are, or are contained in the leaf nodes of the tree data 
structure. In Lagrangian or ``particle'' methods such as SPH, a number of SPH 
particles can be contained in a leaf node or ``bucket''. The maximum number of 
particles per bucket is referred to as $N_B$. In Eulerian or ``grid''-based 
methods the bucket is the smallest grid cell itself, so $N_B$ is effectively 
one. $N$ resolution elements hold radiation intensity values and represent the 
radiation field \acro{} computes. 

Note that although \acro{} has been initially implemented in the SPH code 
\textsc{Gasoline} \citep{wadsleyEt17}, \acro{} is not specific to SPH. The 
method only requires that the simulation volume be hierarchically partitioned 
in space and so it could be used directly in an adaptive mesh refinement code. 
In the case of grid codes, the algorithm is simplified as the final SPH 
particle ray tracing step is not needed.
 
\begin{figure*}
\includegraphics[width=1\linewidth]{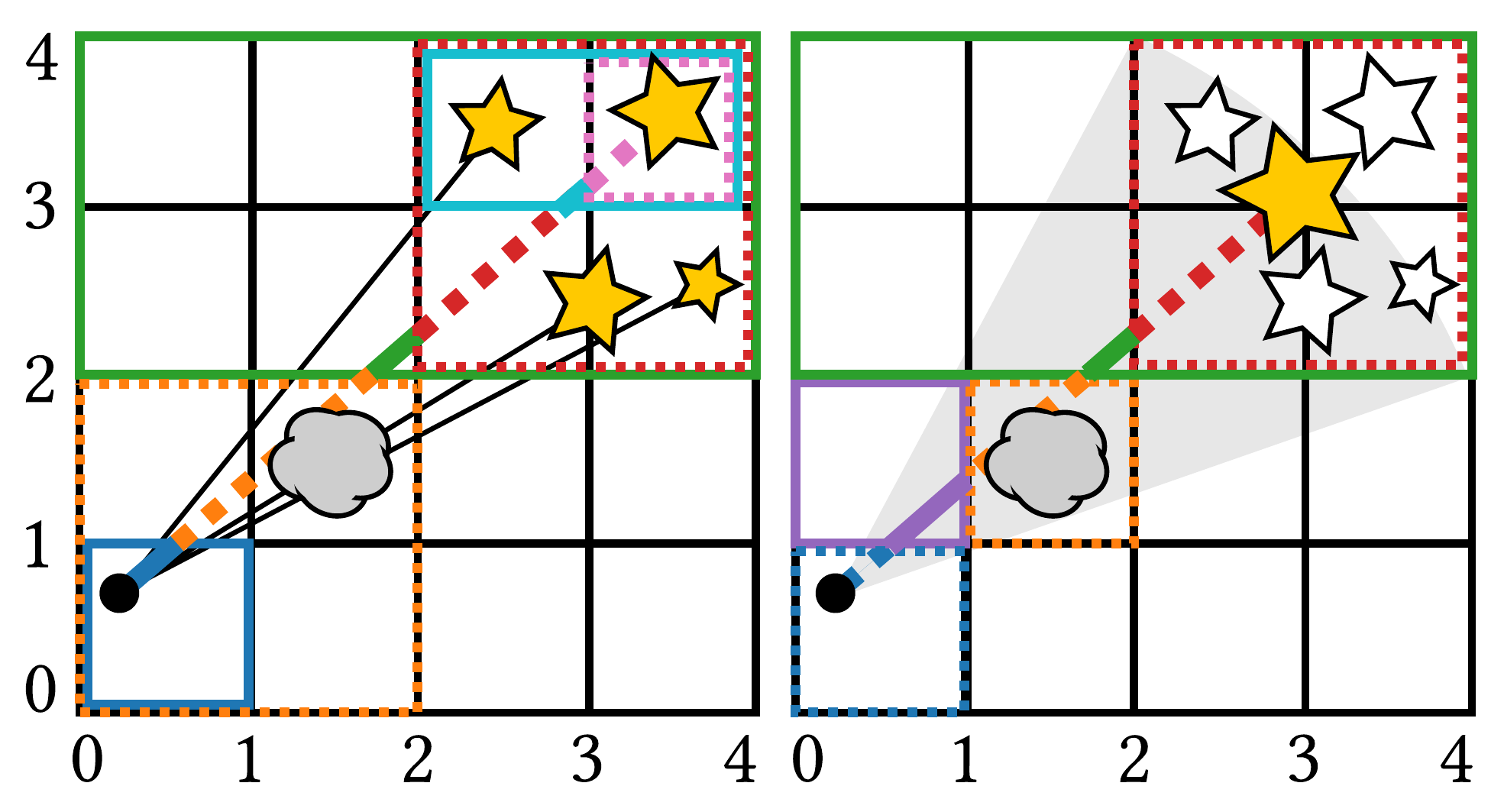}
\caption{A schematic of \acro{} without (left) and with source merging and 
adaptive refinement (right). Coloured ray segments correspond to tree cells 
whose average properties are used to compute the optical depth along that ray 
segment. Dashed and solid lines distinguish consecutive line segments to 
help associate them with their corresponding tree cell. The grey cloud 
represents a feature in the medium that requires refinement in order to be 
resolved. The smaller stars which are yellow in the left panel and white in 
the right panel represent individual radiation sources. The larger yellow star 
in the right panel represents a merged source, as the dashed red cell 
encapsulating all sources meets the opening angle (grey region in right panel) 
criterion.} 
\label{fig:algorithm}
\end{figure*}

\subsubsection{Source Merging}
As mentioned in the introduction, a na\"ive algorithm would compute 
interactions between a resolution element and  all sources of radiation. If we 
assume the number of resolution elements is equal to the number of sources, 
an infeasible number of interactions would need to be computed, with scaling
$\bigO{N^2}$. To mitigate this $N^2$ scaling \acro{} employs source merging 
similar to particle merging in the \cite{barnesHut86} tree-based gravity 
solver which has remained popular in astrophysics 
\citep{benz88,vineSigurdsson98,springelEt01,wadsleyEt04,hubberEt11}. We first
applied radiation source merging in a rudimentary version of \acro{} 
that did not consider extinction of any kind \citep{KannanEt14}.

For a given sink point, sources of 
radiation inside a tree cell are merged together at their centre of luminosity if they meet an 
``opening angle'' criterion. This criterion is defined as 
\begin{equation}
\label{eqn:openangle}
\tO > l/r,
\end{equation}
where $l$ is the distance from the centre of luminosity to the furthest part of
the tree cell, $r$ is distance from the sink to the closest cell edge 
and $\tO$ is the opening angle, a fixed accuracy parameter.  This is equivalent to
the criterion used for gravity in \citet{wadsleyEt04} and ensures parent cells
of a point are always opened.  
Source merging considerably reduces the number of 
interactions \acro{} computes. This is illustrated in the left panel of 
Figure~\ref{fig:algorithm}, where the grey angle represents a cell whose 
angular size meets the opening angle criterion.

The cost savings of source merging can be quantified by integrating the number 
of tree cells that pass the opening angle criterion and whose contents are treated as a single source.
We will call the total count of the cells used $N_{\rm cell}$. We can
estimate $N_{\rm cell}$ by integrating spherical shells of 
thickness $dr$ along the path from a resolution element $r$, and then dividing 
the sphere volume by the volume of the cell, $V_{\rm cell}(r)$.
\begin{equation}
\label{eqn:nsint}
N_{\rm cell} = \int_{R_B}^R \frac{4\pi r^2}{V_{\rm cell}(r)} dr
\end{equation}
The bounds of the integral are $R_B$, the size of a bucket 
cell, and $R$, the length of the simulation volume. Because the number of 
particles in a simulation is proportional to the simulation volume, the 
lower integration limit can be expressed using particle numbers via,
\begin{equation}
\label{eqn:ratio}
\frac{R_B}{R} = \left(\frac{N_B}{N}\right)^{1/3},
\end{equation} 
the cube root of the ratio of the average number of particles per bucket, 
$N_B$, to the total number of simulation particles. Again, note that $N_B$ is 
only needed for particle methods and is one otherwise. The cell volume 
can also be rewritten by cubing the opening angle parameter
\begin{equation}
\label{eqn:vcell}
V_{\rm cell}(r) = l^3 = \tO^3 r^3.
\end{equation}
Substituting gives us the following integral and its solution,
\begin{equation}
\label{eqn:nssoln}
N_{\rm cell} = \int_{(N_B/N)^{1/3}}^R  \frac{4\pi}{\tO^3\,r} dr
\sim \log{N/N_B}.
\end{equation}
This result means that the number of interactions scales like 
$\bigO{\NK \log N}$. This is also the total cost scaling in the optically 
thin regime, as expected given that the RT problem is almost identical to 
the gravity problem in the absence of intervening material.

We next consider ray tracing in the optically thick regime.

\subsubsection{Tracing Rays}
In the presence of absorbing material along a ray, the optical depth needs to 
be computed following
Equation~\ref{eqn:tauint}. To solve this integral numerically, we traverse 
the tree between the source and resolution element to sum the optical 
depth. This requires that the tree partitions and fills space, thus all 
the intervening material is contained in the tree we traverse. Making 
use of properties computed during the tree build, we can compute the optical 
depth of the $i$-th piece of the ray, $\tau_i$, using the intersection length 
of the cell and ray, $s_i$, as well as the average density, $\bar{\rho}_i$, 
and average opacity, $\bar{\kappa}_i$, in the cell
\begin{equation}
\label{eqn:taui}
\tau_i = \bar{\rho}_i \bar{\kappa}_i s_i.
\end{equation}
The total optical depth is then summed up during the tree walk,
\begin{equation}
\label{eqn:tausum}
\tau = \sum_i \tau_i,
\end{equation}
giving us everything needed to evaluate Equation~\ref{eqn:simpflux}. 

This process is illustrated in the left panel of Figure~\ref{fig:algorithm}. 
In this figure ray segments and corresponding cells share 
the same colour. When referring to specific cell colours, they will also be 
identified by two sets of points, in the form $[(x,y),(x,y)]$, corresponding 
to the bottom left and top right vertices of the cell respectively. Dotted 
lines are used to distinguish consecutive ray segments and help associate ray 
segments with their corresponding cells. In the left panel of 
Figure~\ref{fig:algorithm} there are two important things to note. First, 
since we are performing a reverse ray trace, the resolution element denoted by 
the black circle is intrinsically well resolved at the bucket cell (the blue 
cell at $[(0,0),(1,1)]$) level. However, the second point is that as the tree 
is walked upwards, space becomes less resolved. It should be apparent that the 
central parts of the ray are less resolved (the green cell at $[(0,2),(4,4)]$) 
and as one moves towards the source or resolution element the ray becomes more 
resolved (the red cell at $[(2,2),(4,4)]$ and the orange cell 
$[(0,0),(2,2)]$). This can be considered in two ways. If the medium is 
uniform, the algorithm can be extremely efficient while still being able to 
resolve a sharp feature in the radiation field such as an ionization front. 
However, if the medium is highly irregular along the ray the algorithm will 
not be able to resolve sharp density and opacity gradients which could 
significantly alter the optical depth. Thus adaptive refinement is needed 
during the tree walk to accurately calculate the optical depth along the ray.

\subsubsection{Adaptive Refinement}\label{sec:adref}

For each ray, \acro{} decides whether to use the full resolution 
available or if a set of longer ray segments intersecting lower resolution parent 
cells would be sufficient. In principle, the resolution elements themselves 
could be subdivided based on properties associated with RT for even higher 
resolution, but that is beyond the scope of the present work.

Consider the right panel in Figure~\ref{fig:algorithm}. A dense blob of gas 
to be resolved resides in the orange highlighted cell at $[(1,1),(2,2)]$. At 
the point in the tree walk where we reach the orange highlighted cell at 
$[(0,0),(2,2)]$ in the left panel, a decision needs to be made on whether the 
current cell sufficiently represents the medium. This decision is based on a 
refinement criterion. If the cell passes the criterion to refine, rather than 
using its average properties, we recursively check the cell's children until 
the criterion fails. This process ensures we compute a better resolved section 
of the ray. 

Difficulty comes in choosing a refinement criterion that is both accurate and 
efficient. Ideally, refinement occurs when the average optical depth 
in a region may not accurately reflect the true distribution, such as a clumpy 
medium where the average density and opacity is much higher than the 
``effective'' density and opacity \citep{varosiDwek99, hegmanKegel03}. For 
this reason, we developed a new, optical depth-based refinement criterion 
for \acro{}.

\begin{figure}
\includegraphics[width=1\linewidth]{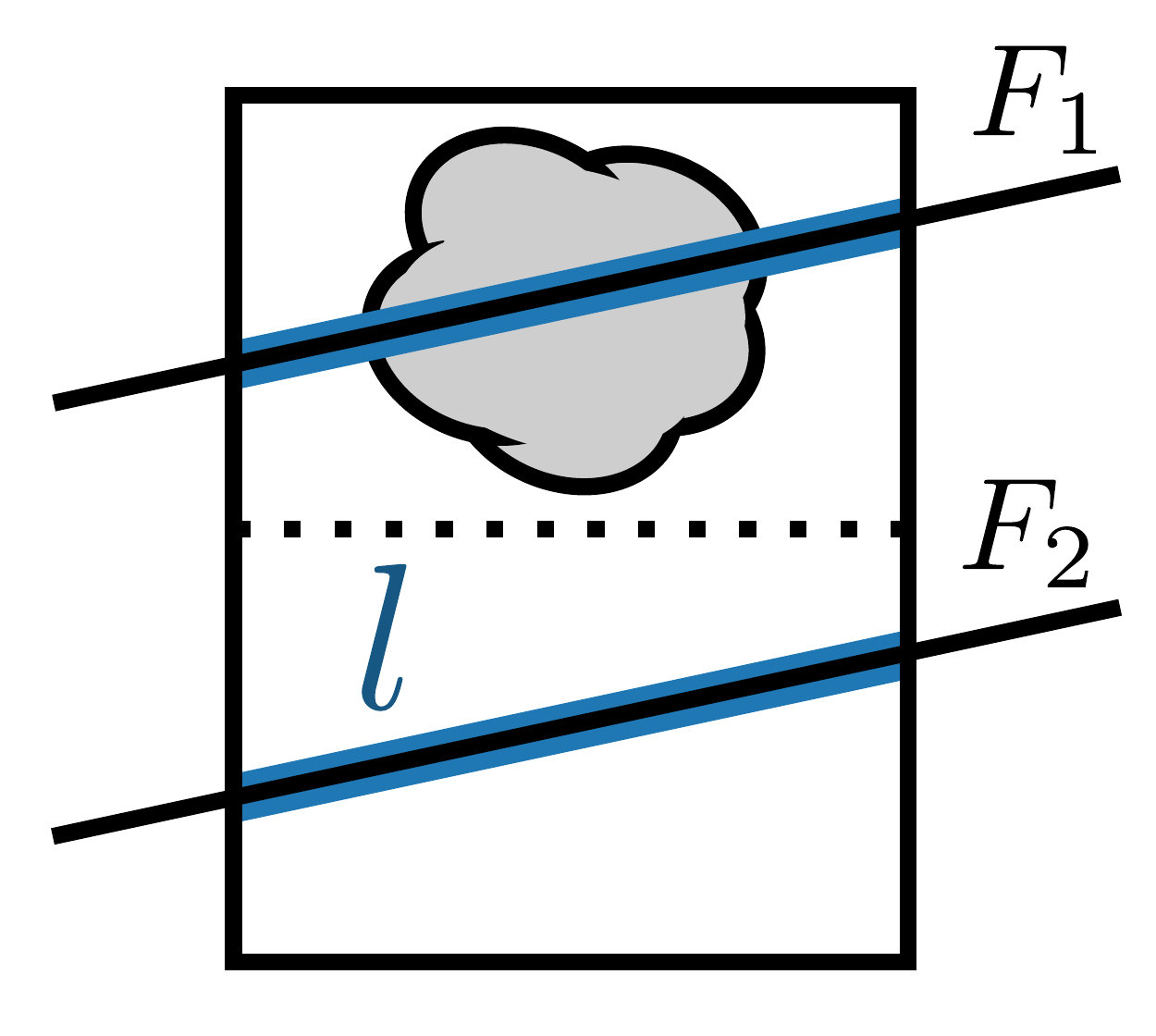}
\caption{Schematic of a cell to be refined. A parent cell intersected by a ray 
contains a feature (grey cloud) to be resolved. The black dotted line 
partitions the parent cell into it's children. The black intersecting rays 
represent the hypothetical case where only a child cell is intersected by a 
ray. The blue outlined sections on each ray correspond to the intersection 
length, $l$,  used to compute the optical depth through each child cell.}
\label{fig:refine}
\end{figure}
Our criterion requires minimum and maximum 
absorption coefficients, $\alpha_{\rm min}$ and $\alpha_{\rm max}$, for each cell.
These are estimated for the three Cartesian directions ($x,y,z$) separately.
Leaf cells are assumed to have a single value $\alpha = \kappa \rho$.
Then, for example,
we estimate the minimum, $x$-direction optical depth of the parent cell via
the minimum $x$-direction optical depths of the child cells.  This requires
taking the minimum for cases where the ray would intersect
one child cell or the other or a sum for cases where the ray passes
through both child cells.  With axis-aligned cells and using 
Cartesian ray directions we know which case applies. We then divide by
the new total $x$-cell width to recover an $\alpha_{\rm min}$ in the $x$-direction for the
parent cell.  The use of maxima and minima effectively takes into account diagonal
rays with similar directions to the Cartesian direction 
(albeit in a conservative fashion because it assumes that two straight rays
corresponding to minimal and maximal optical depth values always actually
exist).  

Proceeding in a bottom-up fashion 
during the tree build, we estimate directional minima and maxima $\alpha$ values for all cells.
 We then take the minima and maxima
over the three directions and save just one $\alpha_{\rm min}$ and one $\alpha_{\rm max}$
for each cell.  

To use the cell-averaged absorption coefficient, $\alpha$, for a ray segment,
we require that substructure within the cell cannot change the final flux
beyond a specified tolerance. This is equivalent to showing that two rays 
intersecting the cell, as in Figure~\ref{fig:refine}, give sufficiently 
similar results. Given $\alpha_{\rm min}$ and $\alpha_{\rm max}$ for that cell 
we can multiply by the ray segment length intersecting the cell, $l$, to 
estimate the minimum, $\tau_{\rm min}$, and maximum, $\tau_{\rm max}$, 
possible optical depths that rays might experience. We can then test the 
following refinement criterion
\begin{equation}
\label{eqn:refcrit}
\tr < \tau_{\rm max} - \tau_{\rm min},
\end{equation}
where $\tr$ is a given, small, tolerance value, and refine if it is true. The 
fractional error in flux, per ray segment, for a given value of $\tr$ is
\begin{equation}
\label{eqn:reffrac}
\frac{F_1 - F_2}{F_1} \leq 1 - e^{-(\tau_{\rm max}-\tau_{\rm min})} 
\lesssim \tr,
\end{equation}
for small $\tr$, making the refinement criterion a convenient choice of 
parameter for controlling error. Figure~\ref{fig:cellplot} is an example of 
\acro{}'s adaptive refinement in action.

It should be noted that if the three Cartesian direction approach were
applied directly to a large cell in isolation, pathological configurations such
as thin planes or filaments not aligned with the axes might be missed.
However, because the maxima and minima are built up from the maxima
and minima of the child cells all the way down to the resolution
scale, any variations on the resolution scale are 
correctly detected by the criterion.  Thus the effective thickness of 
structures is set by the smallest cell size.  In this implementation
that size is comparable to the SPH smoothing length.

For a particle code, if refinement is required at the bucket level, individual 
particles within a bucket must be considered.  A straight forward ray tracing 
scheme similar to SPHRay \citep{altayEt08} can be performed locally on bucket 
particles and their neighbours. This particle-particle step is $\bigO{\NK}$ as 
each particle element interacts with a fixed number of neighbour particles.

Fully characterizing the computational cost of the algorithm, including the 
addition of adaptive refinement, follows the procedure used earlier. Now, 
however, instead of integrating the number of sources we integrate the total 
number of ray segments computed. We will look at two cases, not refining at 
all and fully refining down to the bucket level. This will give us upper and 
lower bound for the algorithm's scaling.

First let's consider the case where the refinement criterion always triggers 
and all rays are resolved down to the bucket level. The number of segments per 
ray is then just the length of a ray divided by the size of a bucket. We can 
express this as,
\begin{equation}
\label{eqn:nseg}
N_{\rm seg} = \frac{r}{R_B} = \frac{r}{R}\left(\frac{N}{N_B}\right)^\frac{1}{3}
\end{equation}
after substituting for $R_B$ using Equation~\ref{eqn:ratio}. Since $\NS$ is 
also the number of rays computed per resolution element, to get the total number of ray segments we
multiply the integrand of Equation~\ref{eqn:nsint} by the number of ray 
segments,
\begin{equation}
\label{eqn:nsegint}
N_{\rm seg} = \int_{(N_B/N)^{1/3}}^R 
\frac{4\pi}{\tO^3}
\frac{1}{R}\left(\frac{N}{N_B}\right)^\frac{1}{3} dr
\sim (N/N_B)^\frac{1}{3}.
\end{equation}
The result is that the total cost of the algorithm scales as
$\bigO{\NK N^{1/3}}$ in the worst-case.

In the case where the refinement criterion never triggers, the ray is split into 
segments made up of the cells traversed in the tree walk of the sub-tree going 
from source to resolution element. The number of cells traversed in a tree walk
is equal to the logarithm of the number of leaf nodes contained within the 
sub-tree. The number of leaf nodes in the sub-tree is also given by 
Equation~\ref{eqn:nseg}, so by taking the logarithm of 
Equation~\ref{eqn:nseg}, we arrive at:
\begin{equation}
\label{eqn:nseg2}
N_{\rm seg} = \log_2\left[\frac{r}{R}\left(
\frac{N}{N_B}\right)^\frac{1}{3}\right],
\end{equation}
where the logarithm is base two, as \textsc{Gasoline} and thus \acro{} is 
implemented using a binary tree. As before, we multiply 
Equation~\ref{eqn:nsint} by the number of ray segments and integrate the 
following:
\begin{eqnarray}
\label{eqn:nsegint2}
N_{\rm seg} & = & \int_{(N_B/N)^{1/3}}^R 
\frac{4\pi}{\tO^3\, r}
\log_2\left[\frac{r}{R}\left(
\frac{N}{N_B}\right)^\frac{1}{3}\right] dr \nonumber\\
& \sim & \log^2(128 N/N_B).
\end{eqnarray}
Thus, in the best-case, the total cost of the algorithm scales as  
$\bigO{\NK \log^2N}$.

\subsubsection{Background Radiation}
In order to treat cosmological simulations properly, we must account for the 
radiation 
coming from the rest of the universe outside the simulation volume. Most 
current codes apply a constant UV field to the entire box, essentially the 
lowest order approximation possible. Some specialized codes like URCHIN 
\citep{altayTheuns13} do a reverse ray trace to the edge of the box, from 
where the background flux is assumed to originate. Others, such as TRAPHIC 
\citep{pawlikSchaye08} allow their ray trace to be periodic.
The cosmic UV radiation field originates from very large distances on the 
order of 100's of Mpc. Thus, for smaller simulation boxes the radiation field 
may be too local.

Instead, we have implemented a method involving tracing ``background sources'' 
similar to URCHIN. ``Background'' particles are distributed in a spiral 
pattern on the surface of a sphere, at the edge of the simulation volume. The number of 
sources can be varied to match the required angular resolution of the 
background. Finding the flux at the centre of a sphere of sources is a problem 
akin to Newton's Shell Theorem. However, because the intensity does not cancel 
like force the solution differs and is as follows:
\begin{equation}
\label{eqn:cosmof}
F(r) = \frac{L}{8\pi R} \ln \left(\frac{R+r}{R-r}\right),
\end{equation}
where $L$ is the total luminosity of the emitting shell, $R$ is the radius of 
the sphere and $r$ is the radius at which the flux is being computed. The 
shape of the function can be seen in Figure~\ref{fig:cosmof} where we have 
plotted the flux as a function of radius for a homogeneous, optically thin 
test volume.
\begin{figure}
\includegraphics[width=1\linewidth]{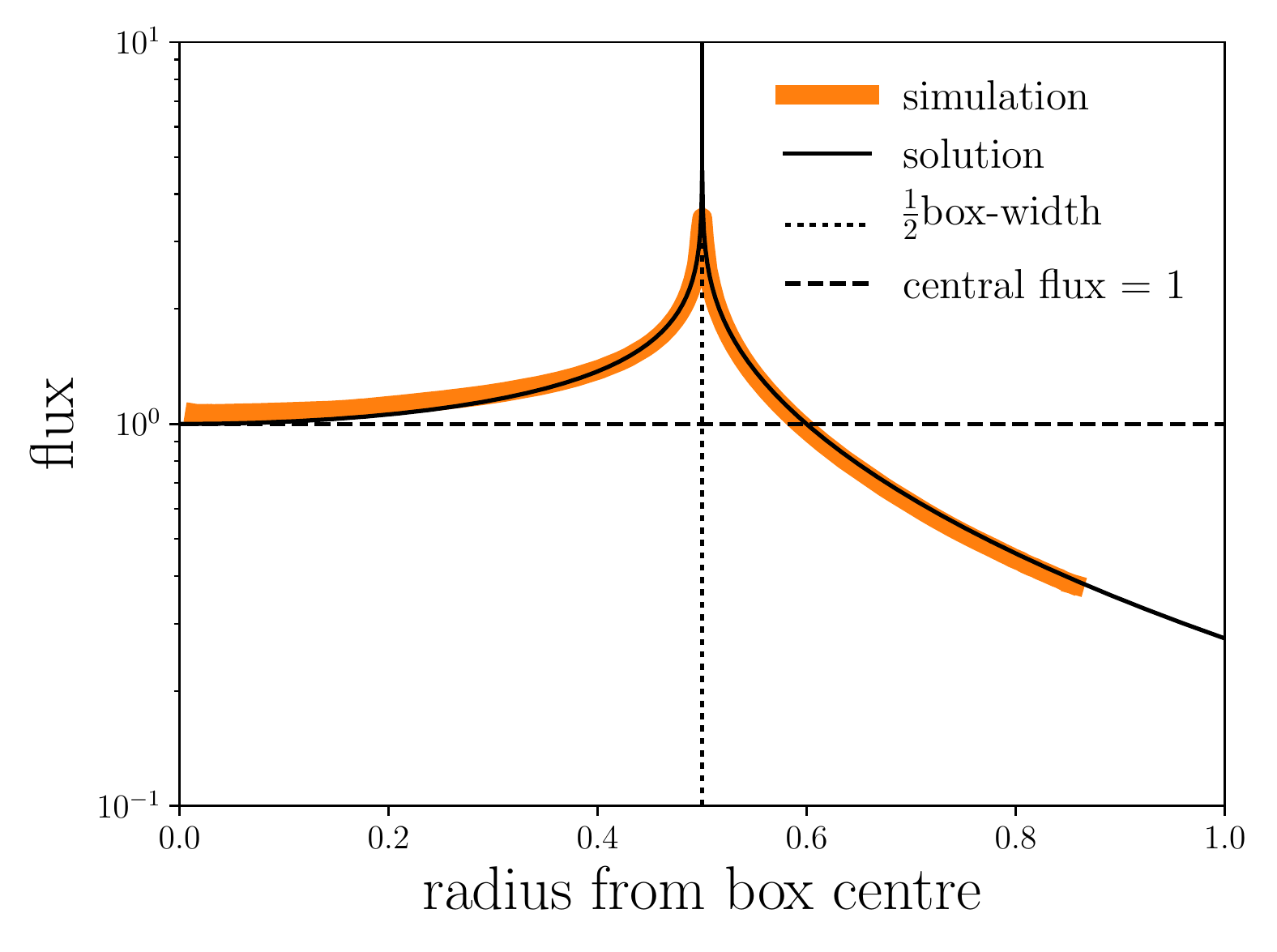}
\caption{Flux as a function of radius from an emitting sphere. \acro{}'s 
solution for background source particles distributed in a spiral on a sphere 
of radius $R=0.5$ (black dotted line) are plotted as the thick orange line. 
The analytical solution given by Equation~\ref{eqn:cosmof} is plotted as a 
thinner black line on top of the numerical solution. A constant flux of one is 
plotted by the black dashed line and is achieved in the inner most region of 
this simulation at around $R \lesssim 0.05$.}
\label{fig:cosmof}
\end{figure}

Note that due to the logarithm in Equation~\ref{eqn:cosmof}, the flux is 
nearly constant at small radii. Since most cosmological zoom in simulations 
only consider gas at a fairly small radius, this setup of background sources 
is an acceptable method of imposing a background flux. A benefit of this 
method is that we can use all of the existing machinery already described, and 
only have to add background star particles as the source of the 
background radiation. Also note that the simulation flux is over estimated 
near the shell.  When merging 
background sources which are all located on the surface of a sphere, the 
merged centre of luminosity will always be at a smaller radius than the sphere 
radius. This can be remedied by forcing merged background sources to always be 
located on the sphere. 

\subsection{Implementation Specifics}\label{sec:specs}
As mentioned earlier, \acro{} is not specific to either \textsc{Gasoline} or 
SPH.  However, in this subsection we 
introduce \textsc{Gasoline} and the specifics of \acro{}'s implementation in 
\textsc{Gasoline}.

\textsc{Gasoline} is a parallel smoothed particle hydrodynamics code for 
computing hydrodynamics and self-gravity in astrophysics simulations. It 
employs a spatial binary tree that is built by recursively bisecting the 
longest axis of each cell. In the current version of \acro{}, a separate tree 
is built for computing radiative transfer. For development purposes, this is a 
convenient choice, but adds extra cost and in the future a single tree
should be adopted, particularly as tree building becomes a significant cost
with adaptive timestepping. A special requirement for the radiation tree is 
that it fills all space to correctly estimate absorption due to ray 
intersections. Both the gravity and hydrodynamics trees squeeze cell bounds to 
the furthest extents of particles within the cell to optimize intersection 
tests which creates gaps between cells.

In the regular tree building phase, \textsc{Gasoline} assigns an ``opening 
radius'' about a cell's centre of mass to each cell in the tree. This radius is
\begin{equation}
r_{\rm open} = \frac{2 B_{\rm max}}{\sqrt{3}\, \tO},
\end{equation}
where $B_{\rm max}$ is the  distance from the centre of mass of particles 
within the cell to the furthest particle from the centre of mass. However, 
since we are using space-filling cells for the radiation tree, it is necessary 
to define $B_{\rm max}$ instead as the distance to the furthest vertex of the 
cell.

The initial method used to compute cell densities during the tree build 
process was to divide the sum of masses of particles within the cell by the 
cell volume, 
\begin{equation}
\rho_{\rm cell} = \frac{\sum_i m_i}{V_{\rm cell}}.
\end{equation}
However, during testing at high levels of refinement we found that the error 
began to increase slightly with increasing refinement accuracy beyond a 
certain level. This was because when refining down to the bucket level, 
$N_{\rm B}=10$ was small enough to introduce Poisson noise in the 
density estimate.  This propagated as errors in the computed radiation field, 
noticeable as noise in otherwise uniform density distributions. To remedy this 
for particle-based methods, each cell uses a volume-weighted average of the particle  
densities,
\begin{equation}
\rho_{\rm cell} = \frac{\sum_i m_i}{\sum_i{\frac{m_i}{\rho_i}}}.
\end{equation}

As noted in Section~\ref{sec:adref}, when sub-bucket level 
refinement is required a ray trace similar to that of 
SPHRay is performed. In this method, all particles in a cell are projected 
down to the ray, and an impact parameter, $b$, is calculated (See 
Figure~\ref{fig:raytrace}).
\begin{figure}
\includegraphics[width=1\linewidth]{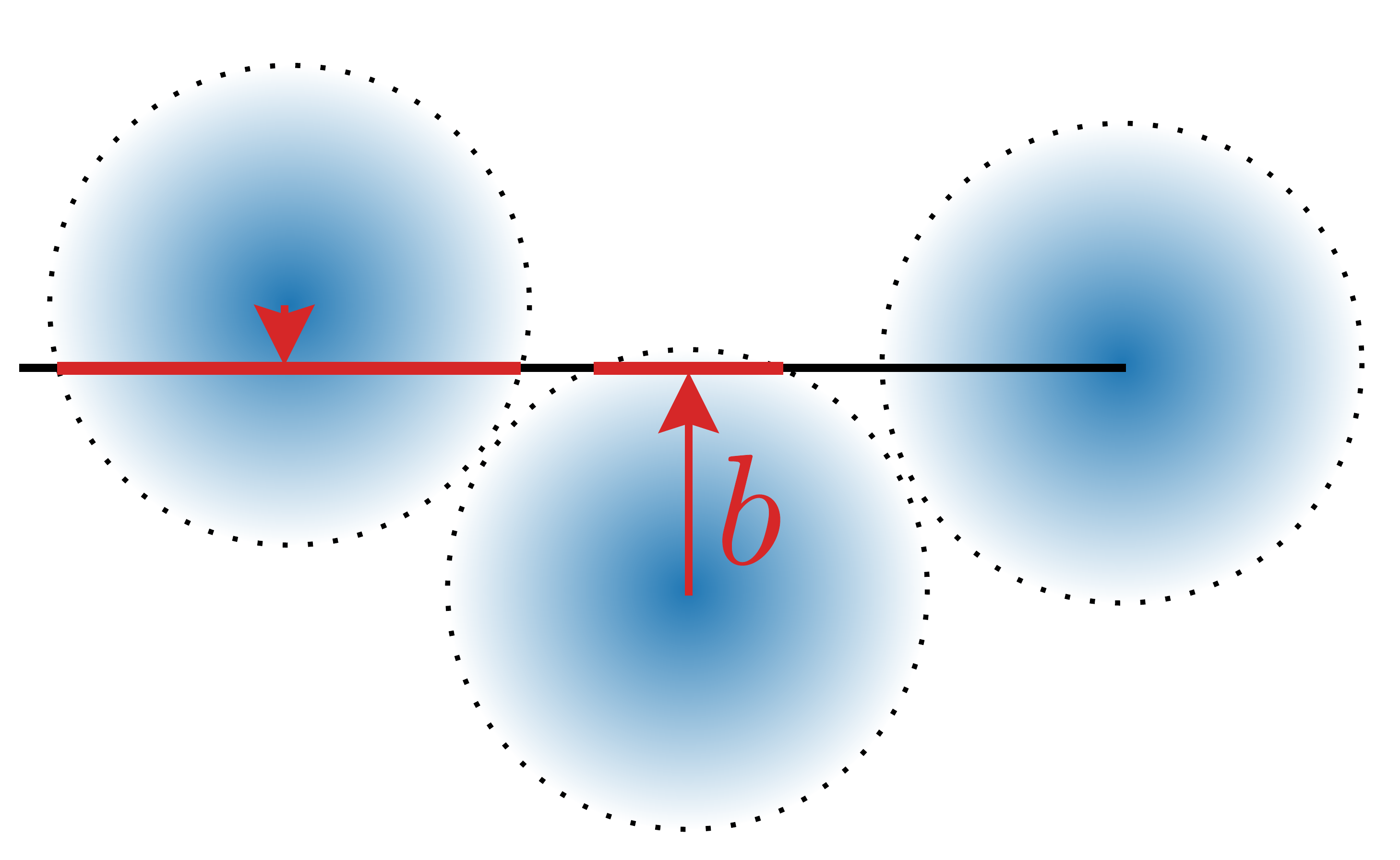}
\caption{The ray tracing scheme employed is similar to that of 
\protect\cite{altayEt08}. In this scheme, the photons are diminished by the 
optical depth along each particle's density field. The receiving particle at 
the termination of the ray does not block photons to itself. Otherwise, the 
front half of the particle would diminish the incoming photons without 
actually having absorbed them.} 
\label{fig:raytrace}
\end{figure}
Since the density field of an SPH particle varies with radius due to the 
smoothed nature of SPH, an (pre-calculated) integral over the smoothing kernel, $W$, must be 
used. Thus, Equation~\ref{eqn:tauint} becomes 
\begin{equation}
\tau_i(s) = \left(m_i\int W\right) \kappa_i ds,
\end{equation}
where $m_i \int W$ represents the effective density along the particular ray 
and $ds$ is the section of the ray intersected by the particle's smoothing 
length (red line segments in Figure~\ref{fig:raytrace}). Note that for the 
receiving particle, its own density field does not contribute to the overall 
optical depth. To see why this must be the case, consider the case where a 
single particle is optically thick. If the front half of the particle 
contributed to absorption, the flux calculated at the centre would be 
effectively zero, and the particle would incorrectly report no heating or 
ionization.  This is essential for correct ionization fronts such as in the 
\strom{} test of Section~\ref{stromgren}.

The implementation operates in parallel in exactly the same way
as the gravity solver, as described in the original \textsc{Gasoline} paper by locally
caching copies of remote tree cells and particles.


\section{Code Tests}\label{sec:tsts}
\subsection{Sinusoidally Perturbed Glass}\label{sinusoidtest}
\subsubsection{Initial Conditions}
To test the accuracy and general scaling of the algorithm we require an 
initial condition (IC) that is representative of a typical use case. For this 
we have created a novel IC comprised of a unit length glass cube of $N$ SPH gas and 
$N$ star particles whose positions have been perturbed by 24 random sinusoidal 
modes. The initial glass of particles is built from copies of the $16^3$ glass 
used to create ICs for other tests of \textsc{Gasoline} 
\citep{wadsleyEt17}. The total mass of gas particles is one, and the opacity 
of each particle is also one. This results in an optical depth across the 
width of the box of $\sim$1, making the simulation volume marginally  optically thin 
overall with dense, optically thick filamentary structure and underdense voids 
qualitatively similar to the cosmic web. Each star particle is assigned a 
luminosity of one. A slice of this density distribution is plotted in 
Figure~\ref{fig:sine_rho}. Appendix \ref{sec:icnd} contains a detailed 
explanation of how this IC was created including a table of modes used.
\begin{figure}
\includegraphics[width=1\linewidth]{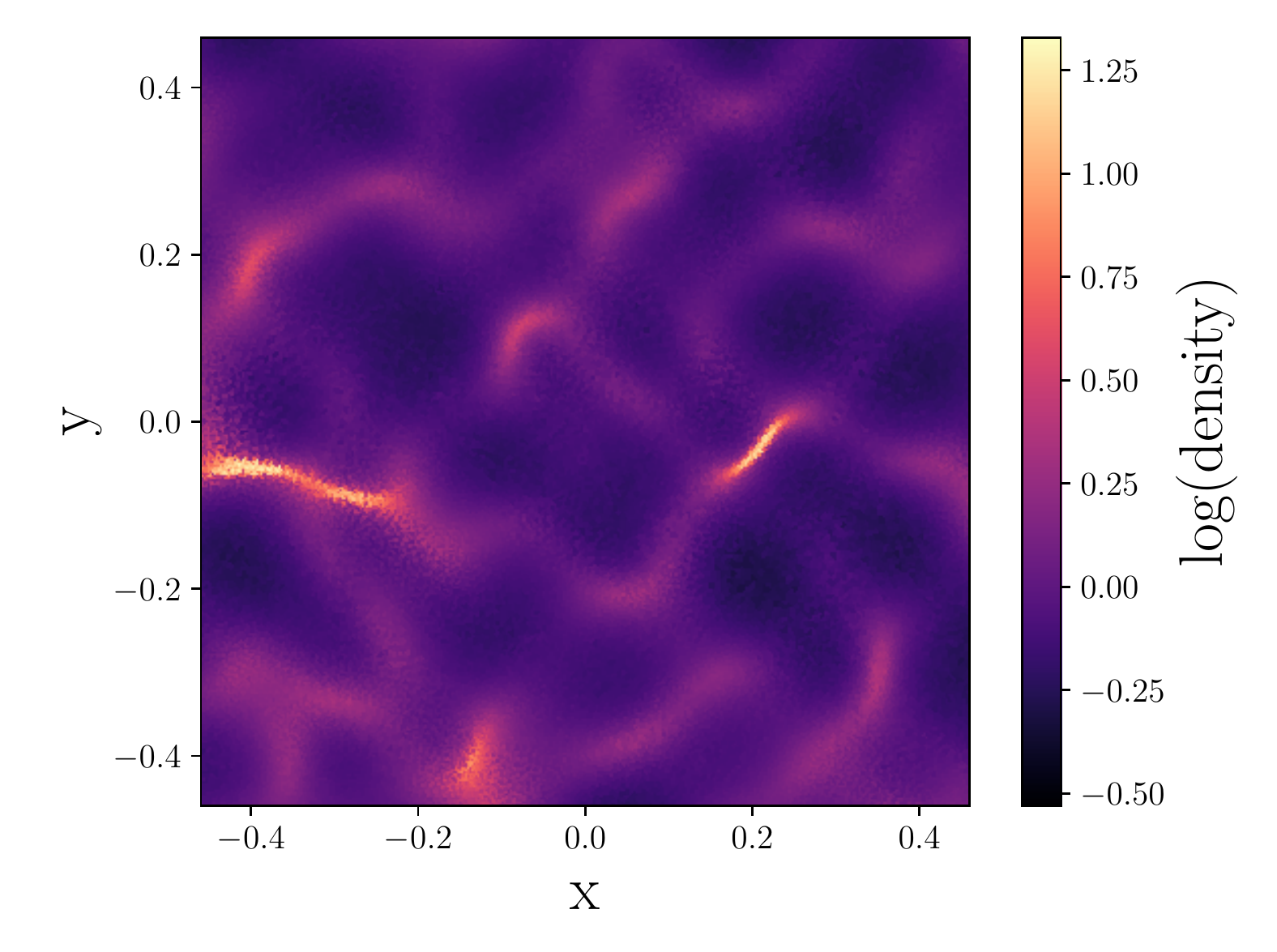}
\caption{A $z$-plane slice of the sinusoidally perturbed glass IC. The optical 
depth along the longest filament in the slice (left, just below $y=0$) is
$\tau\approx 4$. The optical depth across the largest void (above the 
aforementioned filament) is $\tau\approx 0.1$.}
\label{fig:sine_rho}
\end{figure}

\subsubsection{Opening Angle}
The opening angle criterion's effect on accuracy and cost was tested by 
ray tracing the optically thin, sinusoidally perturbed glass IC with $\tO$ 
varying between 0 and 1. The results of this test are plotted in 
Figure~\ref{fig:openangle}. The measure of cost is plotted as the total number 
of rays, $N_{\rm rays}$, computed per resolution element on the left $y$-axis. 
The number of rays is equivalent to the number of radiation sink-source 
interactions computed in a simulation timestep. Using rays as a measure of 
cost allows us to isolate the effects of the opening criterion on cost. On 
the right $y$-axis we have plotted the root mean squared (RMS) fractional 
error relative to the radiation field computed with $\tO=0$. This test was run 
with $\tr = 0.1$ and $N=64^3$ star and gas particles.
\begin{figure}
\includegraphics[width=1\linewidth]{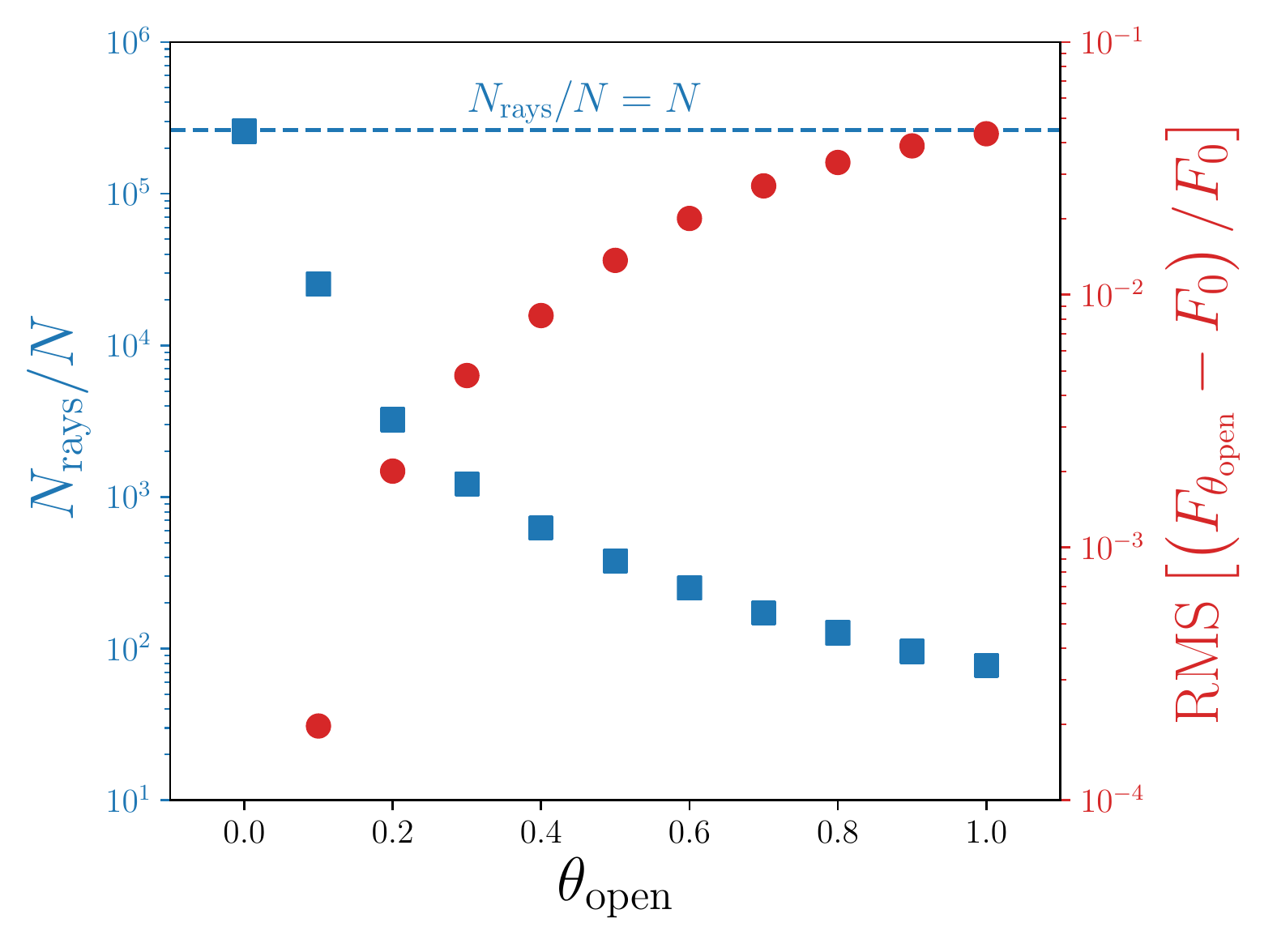}
\caption{A plot of cost and accuracy as a function of opening angle. The 
number of rays computed per resolution element is plotted in blue, on the left 
$y$-axis with square markers. The blue dashed line shows $N_{\rm rays}/N 
= N$ at an opening angle of $\tO=0$, meaning \acro{} can perform an 
$\bigO{N^2}$ ray trace if desired (omitting the cost of absorption). The RMS 
error in flux relative to $\tO=0$ is plotted in red, on the right $y$-axis 
with circular markers.}
\label{fig:openangle}
\end{figure}

At $\tO=0.75$, the value used in all other tests and the 
default value for $\tO$ in many gravity solvers, 200 rays are 
computed per resolution element with an RMS fractional error of 3\%. To 
achieve a RMS fractional error of about 1\%, we suggest that a lower opening 
angle of approximately $\tO=0.45$ should be used. $\tO=0.45$ costs only 500 
rays per resolution element, which is still much less than interacting with 
all $64^3$ ($2.6\e{5}$) sources.

\subsubsection{Refinement Criterion}
Testing the refinement criterion is similar to testing the opening angle 
criterion. Again, the sinusoidally perturbed glass IC was simulated but now 
with a varying $\tr$ value. The results of this test are plotted in 
Figure~\ref{fig:refcrit}. The min and max values for $\tr$ were chosen
to show how the cost curve flattens out on either side: the left hand side being 
where refinement has occurred down to the bucket level and the right hand side 
being where refinement is never done. An opening angle of 0.75 was used and 
$N=64^3$ for both star and gas particles. Cost is plotted on the left $y$-axis 
and RMS fractional error on the right $y$-axis. The measure of cost is now 
the number of ray segments per resolution element, since the refinement 
criterion controls the number of ray segments a single ray is broken up into. 
The measure of accuracy is again the RMS fractional error, but now relative to 
the radiation field computed with $\tr=1\e{-8}$, which ensures maximum
resolution everywhere.
\begin{figure}
\includegraphics[width=1\linewidth]{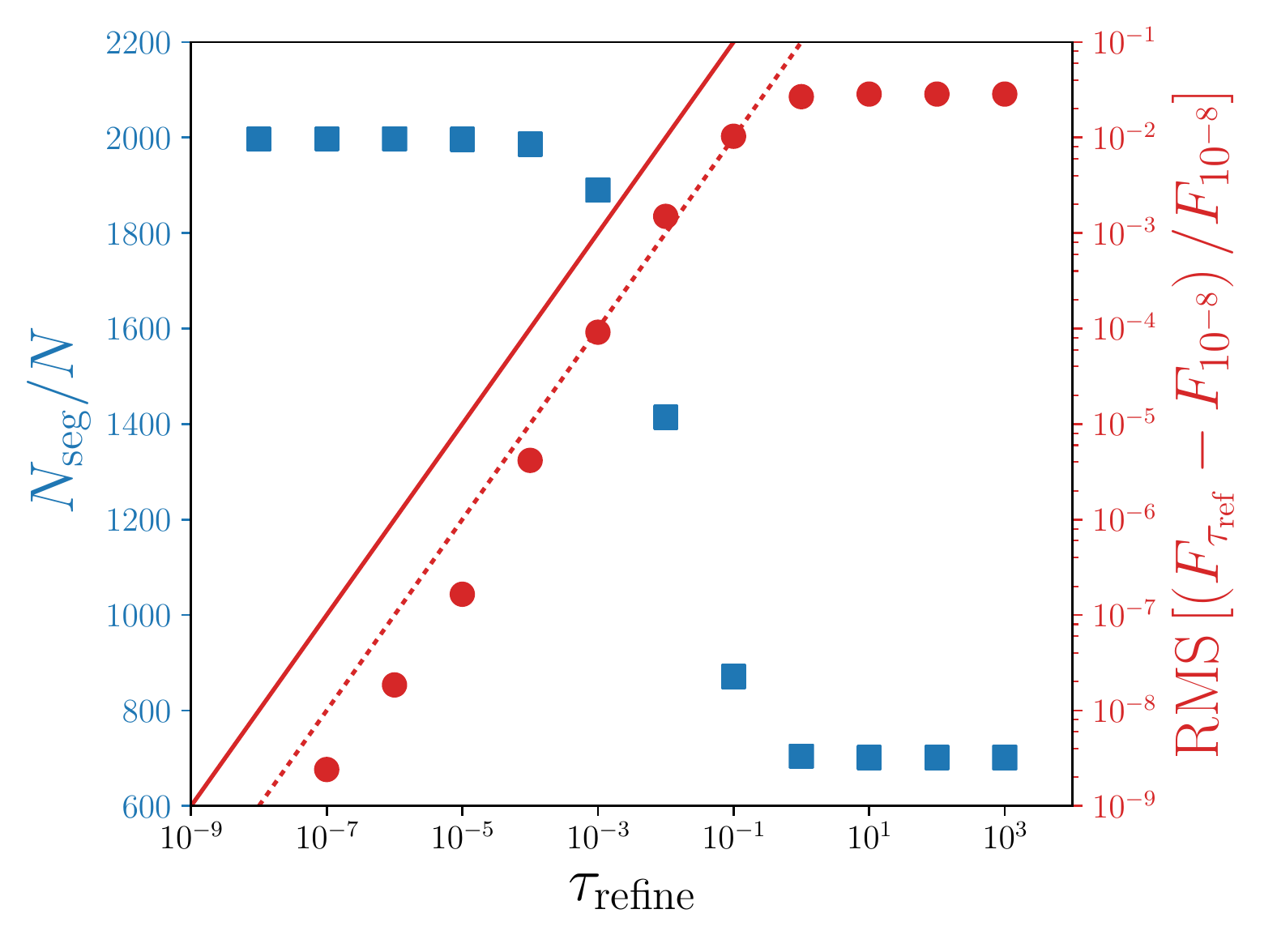}
\caption{Algorithmic cost and accuracy as a function of refinement criterion. 
The number of ray segments computed per resolution element is plotted in blue, 
on the left $y$-axis with square markers. The RMS error in flux relative to 
$\tr=10^{-8}$ is plotted in red, on the right $y$-axis with circular markers. 
Lines of ${\rm error} = \tr$ and ${\rm error} = \tr/10$ are plotted as red 
solid and dotted lines respectively. The upper line represents maximum 
allowable error \textit{per ray segment}. The RMS relative error roughly 
follows the lower line, an order of magnitude lower.}
\label{fig:refcrit}
\end{figure}

At $\tr=0.1$, 1\% RMS fractional error is achieved with a cost of 
approximately 850 ray segments computed per resolution element (including all 
rays), less than half the cost of refining all the way to the bucket level. 
Note also that RMS fractional error as a function of $\tr$ behaves 
predictably, lying below the error = $\tr$ line and roughly following the 
error = $\tr/10$ line plotted in Figure~\ref{fig:refcrit}. This shows that the 
error per ray segment is well controlled by our refinement criterion and 
considerably lower than $\tr$ on average.

The RMS fractional error plateaus at 2-3\% in this test. In this particular 
implementation of \acro{}, the walk along the ray goes up from both the bucket 
where the radiation sink resides and the opened cell where the source resides, 
to the top of the tree. This built in level of refinement is the reason for 
the low maximum error. Other implementations, that walk the ray top down or up 
and then back down the tree, would need to rely more, or solely, on the 
refinement criterion. In principle, such a method could perform better than 
$\bigO{N\log^2 N}$.

\subsubsection{Scaling}\label{scalingtest}
\begin{figure}
\includegraphics[width=1\linewidth]{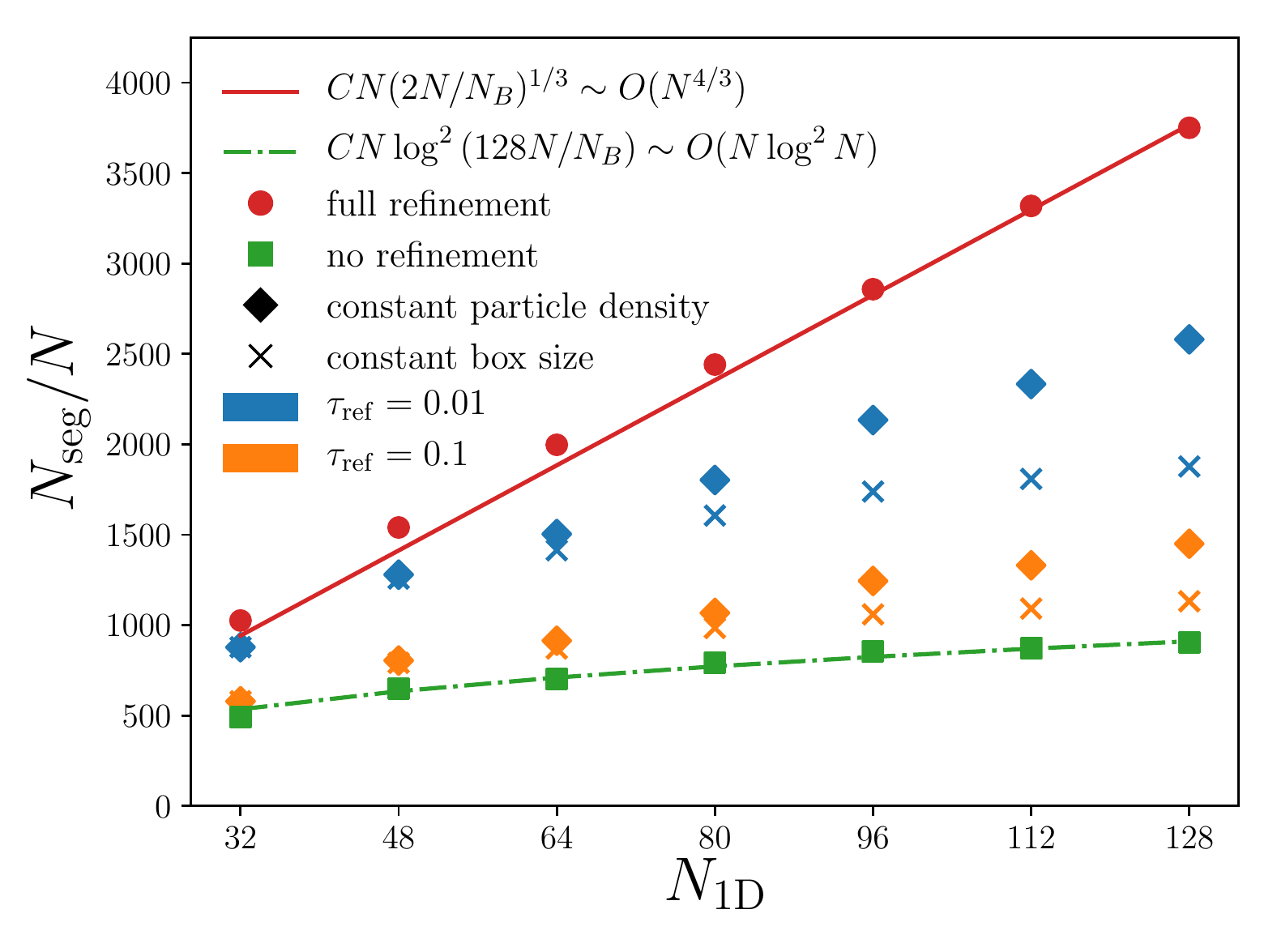}
\caption{Cost, quantified as the number of computed ray segments per 
resolution element, is plotted as a function of $N_{\rm 1D} = N^{1/3}$. 
\acro{}'s theoretical upper and lower scaling bounds are plotted as red 
(solid) and green (dash-dot) lines respectively. The corresponding simulation 
data points are plotted as red circles and green squares. Simulation data 
points intermediate to the scaling bounds are plotted as combinations of two 
parameters - refinement criterion parameter value and the type of scaling. Tests run 
with a refinement criterion of $\tr=0.1$ are coloured orange and make up the 
lower two sets of intermediate data. The upper sets of intermediate data, 
coloured blue, are tests run with $\tr=0.01$. Diamond markers denote weak 
scaling tests and $\times$ shaped markers denote strong scaling tests.}
\label{fig:pscale}
\end{figure}
To test cost scaling as a function of $N$, we hold $\tO$ constant at 0.75 and
vary $N$ between $32^3$ and $128^3$ in steps of $N_{\rm 1D} = 16$ for both gas 
and star particles. To substantiate the best and worst-case theoretical 
scaling claims made in Equations~\ref{eqn:nsegint2} and~\ref{eqn:nsegint} 
respectively, the sinusoidally perturbed glass IC was simulated with 
$\tr=1\e 6$ to ensure refinement was never performed and with $\tr=0$ to 
ensure refinement was always performed down to the bucket level. Data from 
these tests and the fitted theoretical lines are plotted in 
Figure~\ref{fig:pscale} and correspond very closely to each other. Note that 
the only parameter used to fit the theoretical lines is a constant factor 
multiplying Equations~\ref{eqn:nsegint} and~\ref{eqn:nsegint2}.

Scaling behaviour between the upper and lower limits was probed in two ways. 
Firstly, simulations were run with $\tr$ values of 0.1 and 0.01. Secondly, 
strong and weak scaling cases were simulated. The strong scaling case being 
where the simulation volume was held constant and particle number increased. 
This is analogous to increasing the resolution in a standard galaxy 
simulation. The weak scaling case is the opposite, in which the simulation 
volume is increased and particle density is held constant. This is analogous 
to simulating larger and larger cosmological boxes to achieve larger 
statistical samples. Note that the previously described tests of the upper and 
lower scaling bounds were only run as strong scaling tests.

Results from these tests are shown in Figure~\ref{fig:pscale}.  There
are two interesting things to note.  Firstly, with refinement we can
maintain a scaling quite similar to the best case of $N \log^2{N}$ in
this representative test.  Secondly, the strong scaling case, 
which is typically the harder case to scale effectively in other respects 
(e.g. parallelism), scales better than the weak scaling case. The strong 
scaling data is closer to $N\log^2{N}$ and costs less than the weak scaling 
case for the same $N$. This is because the larger boxes in the weak scaling 
case have larger total optical depths and thus require more ray segments to 
achieve the same flux accuracy.  

\subsection{Isothermal Spheres}\label{spheretest}
\subsubsection{Initial Conditions}
The sinusoidally perturbed glass IC tests a generally optically thin, smooth 
density distribution. This is a good proxy for many astrophysical cases of 
interest, such as late stage galaxy evolution. We now show how well \acro{}'s 
refinement criteria can handle compact, optically thick features. We created 
an IC featuring a single radiation source positioned in the top left corner 
and four spheres with $1/r^2$ density profiles (mimicking self-gravitating 
dense objects) embedded in a uniform region, with opacity and density both set 
to one. The four isothermal spheres have a density distribution given by
\begin{equation}
\rho(r) = \frac{\rho_0 \epsilon^2}{r^2 + \epsilon^2},
\end{equation}
where the softening length is $\epsilon=0.002$ and the central density, 
$\rho_0=626$. The IC was made starting with a uniform density glass of fixed 
mass particles. SPH gas particles were added inside the sphere radii. To do 
this the uniform glass was duplicated and associated with negative radii for a 
given sphere. A mapping from this initial space (including both negative and 
positive radii) to positive radii in the final space was calculated 
analytically that gave the desired isothermal density profile for that sphere
while maintaining a glass-like distribution. Any duplicated particles that did 
not map to positive radii were then deleted. This technique is able to embed 
arbitrarily large non-linear density perturbations in any uniform density 
glass.

The chosen parameters set the maximum optical depth through a sphere to 
$\tau_{\rm max}=4$ (a 98\% reduction in flux) and the density at the edge of 
the spheres to one, matching the unit density of the uniform background. The 
isothermal spheres have a radius of 0.05 of the box length and are shown as 
grey circles in Figure~\ref{fig:cellplot}.  
\begin{figure*}
\includegraphics[width=1\linewidth]{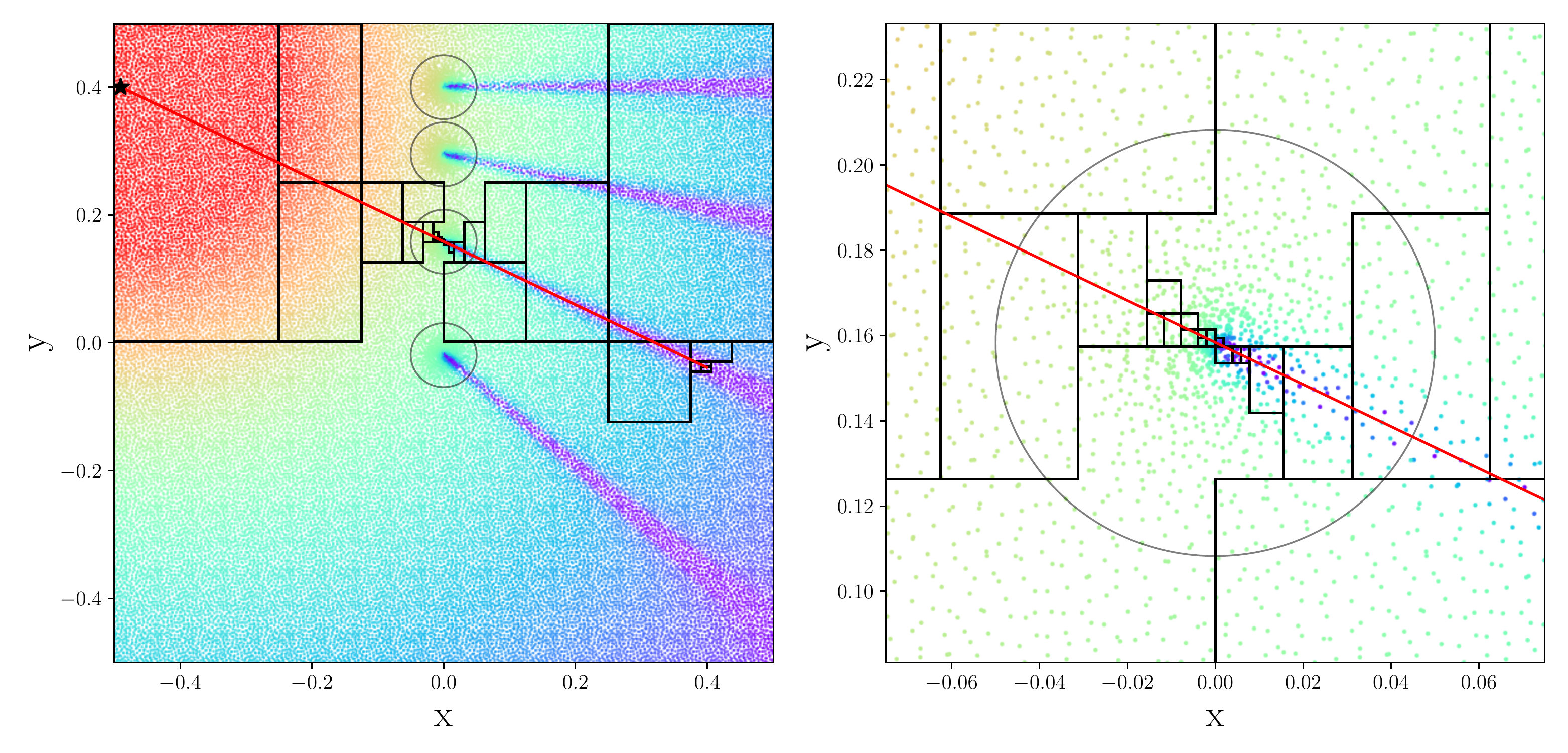}
\caption{Left: \acro{}'s adaptive refinement criterion (at $\tr =0.1$) 
resolving isothermal spheres in a uniform environment. Particles in a slice 
along the $z$-plane of the isothermal spheres IC are coloured by the logarithm 
of their flux value (high - low flux, red - purple). The red line represents 
the ray traced from the radiation source (black star) to the receiving cell. 
Black rectangles represent the spatial boundary of the tree cells used compute 
the optical depth of the intersecting ray segment. Right: A zoom-in of the 
sphere intersected by the ray to focus on the refinement across the sphere 
itself.}
\label{fig:cellplot}
\end{figure*}
The spheres are centred on the $x$ and $z$ axis with $y$ coordinates given by
\begin{equation}
y_i = 0.75 - 1.3^{-(4-i)},
\end{equation}
where $i$ runs from zero to three. The radiation source, denoted by the black 
star in Figure~\ref{fig:cellplot}, is located at $x=0.49$, $y=y_0$ and $z=0$. 
The total number of particles in the IC is $N=4,111,624$.

The spheres produce shadows away from the source. Accurate shadows can only be 
cast if the sharply peaked spheres are resolved correctly by the refinement 
criterion. Errors arising  specifically in the optically thick regime can be 
isolated by looking at particles in shadow.

\subsubsection{Refinement Criterion}
The effects of the refinement criterion on accuracy and cost in this test were 
analyzed similarly to the previous test. The main addition in 
Figure~\ref{fig:isosph} is that the subset of particles in shadow has its RMS 
fractional error plotted separately to highlight the refinement criterion's 
performance in the optically thick regime.
\begin{figure}
\includegraphics[width=1\linewidth]{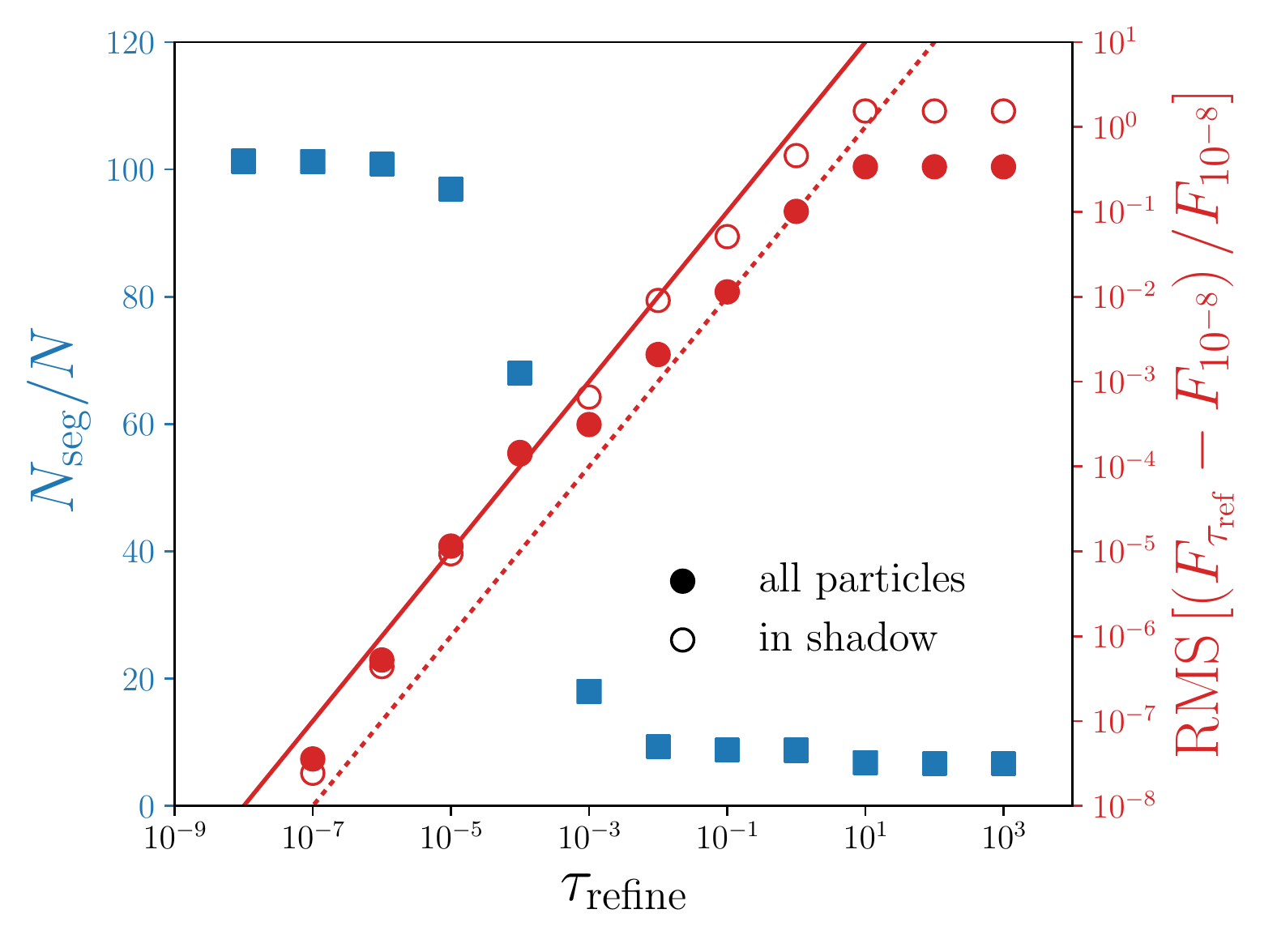}
\caption{A plot of cost and accuracy as a function of refinement criterion. 
The number of ray segments computed per resolution element is plotted in blue, 
on the left $y$-axis with square markers. The RMS error in flux relative to 
$\tr=10^{-8}$ is plotted in red, on the right $y$-axis with circular markers. 
Lines of ${\rm error} = \tr$ and ${\rm error} = \tr/10$ are plotted as red 
solid and dotted lines respectively. Solid circular markers represent RMS 
relative errors computed on all resolution elements and empty markers 
represent only the resolution elements that fall in shadow. The in-shadow 
errors are larger for looser refinement criterion, continuing to follow the 
${\rm error} = \tr$ line.} 
\label{fig:isosph}
\end{figure}
As before, $\tr=0.1$ achieves an overall RMS fractional error of 1\% with very 
little cost. However, when restricting the focus only to those particles in 
shadow, the same refinement parameter produces much higher errors
($\sim$ 8\%). 
Decreasing the refinement 
parameter by an order of magnitude to $\tr=0.01$ predictably decreases the RMS 
fractional error on particles in shadow to 1\%, with a negligible 
increase in the cost from that at $\tr=0.1$ as most of the volume is fairly 
optically thin and does not need to be refined in either case.

The $\rm{error}=\tr$ and $\rm{error}=\tr/10$ lines are again plotted in 
Figure~\ref{fig:isosph}. For the most part the RMS fractional error is 
contained between these lines, with only two of the in-shadow points at 
$\tr=1\e{-5}$ and $1\e{-4}$,  and one of the all-particle points at  
$\tr=1\e{-4}$ sitting marginally above the $\rm{error}=\tr$ line. The error 
bound represented by Equation~\ref{eqn:reffrac} is tighter for this test, with 
the overall error closer to $\tr$.

The isothermal spheres test is an especially difficult test as there is only 
one source, and the errors are more systematic. Thus, the errors are less 
likely to cancel the way random errors often do, for example, with many sources. 
Such random cancellations mean that overall errors for \acro{}
typically perform better than the bound given by Equation~\ref{eqn:reffrac}.

The isothermal spheres test is useful as it is representative of structure
commonly found in astrophysics and because it has an analytic solution. 
However, in SPH it is difficult to represent sharp density gradients with 
discrete resolution elements. This causes in-shadow particle flux errors 
relative to the analytic solution (not shown) to be up to an order of magnitude higher 
than the errors relative to the $\tr = 10^{-8}$ simulation plotted in 
Figure~\ref{fig:isosph}. These errors are associated with the discrete 
representation of the density profile rather than the radiative transfer 
method.

\subsection{\strom{} Sphere Test} \label{stromgren}
\subsubsection{\strom{} Sphere Theory}
The \strom{} sphere is a theoretical ionized sphere of gas first discussed by 
Bengt \strom{} in 1938 \citep{stromgren39} as a model of the HII region around 
a hot, young star. The theoretical ICs consist of a uniform 
density cloud of neutral hydrogen gas with an ionizing source of radiation at 
its centre. As photons from the source ionize the hydrogen, the optical depth 
of the gas decreases and so the ionizing photons are able to travel further 
and further from the source creating a moving ionization front.  Eventually,
a radius is reached such that the total ionization rate equals the 
recombination rate.  At this point, the front reaches an equilibrium, creating 
a stable sphere of ionized hydrogen. The \strom{} sphere test has become a 
common code test in RT methods papers \citep{pawlikSchaye08, pawlikSchaye11, 
petkovaSpringel11} and comparison papers \citep{ilievEt06, ilievEt09}, as it 
is a simple test of a method's ability to resolve ionization fronts and 
achieve equilibrium behaviour that may be compared with analytic results.

The equilibrium radius or \strom{} radius, $R_S$, is the radius at which the 
ionization and recombination rates are equal \citep[e.g.][]{tielens05},
\begin{equation}
R_S = \left(\frac{3}{4\pi}\frac{\dot{N}_\gamma}{\alpha\, n^2_{H}}\right)^{1/3},
\end{equation}
where $\dot{N}_\gamma$ is the source luminosity in photons per second, 
$\alpha$ is the recombination rate and $n_H$ is the hydrogen number density. 
One can also solve for the radius as a function of time 
\citep[e.g.][]{spitzer78},
\begin{equation}\label{eqn:stromtime}
R(t) = R_S\left[1-\exp\left(t/t_{\rm rec}\right)\right]^{1/3}
\end{equation}
where $t_{\rm rec = 1/{n_H \alpha}}$ is the recombination time of the 
gas. The above derivation assumes a ``sharp'' ionization front, meaning the 
transition from ionized to neutral hydrogen occurs across an infinitesimally 
thin region. In practice, there is a finite transition region and structure
interior to the \strom{} radius. 
In order to solve for the non-sharp ionization front we must consider the 
hydrogen ionization equation
\begin{equation}\label{eqn:hion}
\frac{\partial n_{\rm HII}}{\partial t} = c \sigma n_{\rm HI} n_{\gamma} - 
\alpha n_{\rm e} n_{\rm HII},
\end{equation}
where $n_x$ is the number density of species $x$, HI and HII label neutral 
and ionized hydrogen respectively, $\gamma$ denote photons, $\sigma$ is the 
ionization cross-section, $c$ is the speed of light and $\alpha$ is the 
recombination rate. Note that we have omitted collisional ionization in 
Equation~\ref{eqn:hion}, which is customary for this test, however it 
should be included in general. By integrating the ionization equation and the 
flux equation with absorption (Equation~\ref{eqn:simpflux}), we get a solution 
for the relative abundance of HI and HII as a function of radius and as a 
function of time \citep{osterbrockFerland2006}. In the following tests, we 
include both the theoretical sharp front solution and non-sharp front 
solutions from the \cite{ilievEt06} comparison paper to compare to our 
results. We also attempt to duplicate the ICs of 
\cite{ilievEt06} as closely as possible.  It should be noted that a time-step
limit associated with ionization and temperature changes is required to correctly
follow ionization fronts.  We employed the pre-exisiting ionization, heating
and cooling integrator in \textsc{Gasoline} with no changes other than
using the standard coefficient values for these tests described below.

\subsubsection{The Isothermal \strom{} Sphere}
\begin{figure}
\includegraphics[width=1\linewidth]{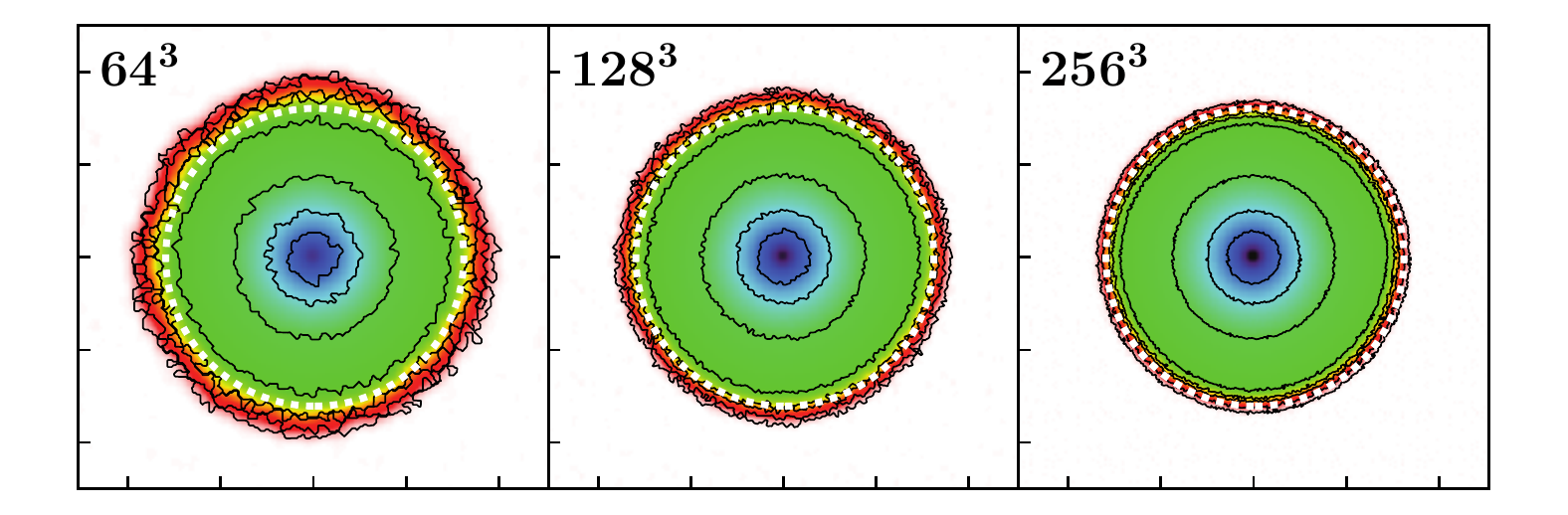}
\includegraphics[width=1\linewidth]{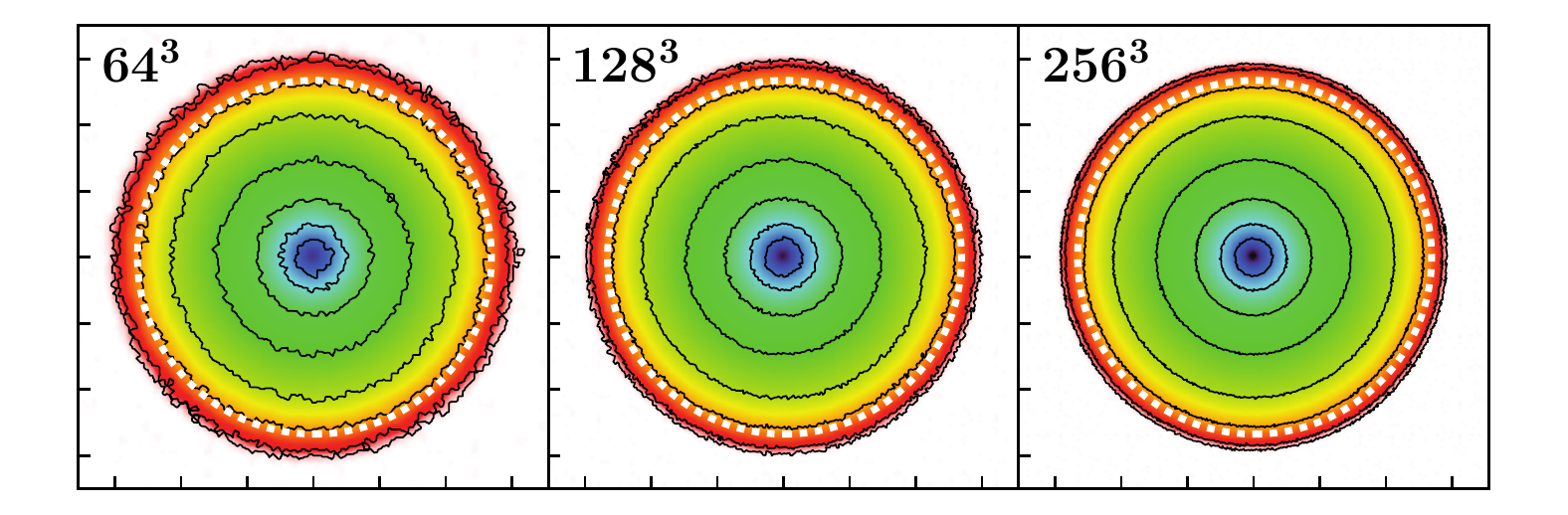}
\caption{A slice through the $z$-plane of the isothermal \strom{} sphere
test at $t=30$ Myr (top row) and $t=500$ Myr (bottom row). Particle 
resolutions increase from left to right denoted by the $N$ value in the top 
left corner of each pane. Axis ticks are spaced 2 kpc apart, so note that 
ionized spheres in the top row are a fraction of the volume and particle 
resolution of spheres in the bottom row. The colormap represents neutral 
fraction, $\rm x$, and is similar to that of \protect\cite{pawlikSchaye08} and 
\protect\cite{pawlikSchaye11} to allow for ease of comparison.  We use the same
contour levels: $\rm x$ = 0.9, 0.5, $\log \rm x$ = -1, -1.5, 
-2, -2.5, -3 and -3.5. The white dashed line is a circle of radius given by 
Equation~\ref{eqn:stromtime}, the sharp, time dependent solution to the 
isothermal \strom{} sphere.}
\label{fig:stromslice}
\end{figure}
In the simplest case, the ionizing source is assumed to emit monochromatic 
photons at 13.6 eV and the gas is held at fixed temperature of $T=10^4$ K.
We refer to this case as the isothermal \strom{} sphere.  The medium is 
initially neutral with a uniform density of $n_{\rm HI} = 10^{-3}$ 
$\rm cm^{-3}$. We use an ionization cross-section of $\sigma = 6.3 \times 
10^{-18}$ $\rm cm^{-2}$ and a recombination rate of $\alpha = 2.59 \times 
10^{-13}$ $\rm cm^{-3}$ $\rm s^{-1}$, typical of $10^4$ K gas. An ionizing 
source is turned on at $t=0$ and emits at a rate of $\dot{N}_\gamma = 5 
\times 10^{48}$ photons $s^{-1}$.  These values yield a \strom{} radius of 
$R_S = 5.38$ kpc and a recombination time of $t_{\rm rec} \approx  125$ Myr. 

We note that \cite{ilievEt06} use a 6.6 kpc cube which only contains a single
octant of the \strom{} sphere for their testing. We have opted to use an 16 
kpc cube, increasing the maximum front radius to 8 kpc to avoid any edge 
effects (the sphere gets close to the edge of the box for some codes in 
the above paper). In order to aid comparison, we still normalize radius values 
to 6.6 kpc, as is done in \cite{ilievEt06}. As well, we have not imposed a 
floor on the HII fraction of 0.001, as was done in their paper. Because the 
resolution used in the \cite{ilievEt06} comparison paper was never 
specifically given, we have opted to run the test with $N = 64^3$, $128^3$ and 
$256^3$ particles to represent the \textit{entire} sphere. These resolutions 
correspond to single octant resolutions of $N=32^3$, $64^3$ and $128^3$ in 
\cite{ilievEt06}. Varying the number of particles also allows us to invesitgate
at how \acro{} converges with resolution. We have run our \strom{} sphere 
tests with fixed accuracy parameters of $\tO = 0.75$, 
$\tau_{\rm ref} = 0.1$.

Figure~\ref{fig:stromslice} is a slice through the $z$-plane of the 
simulation. The colour map shows the neutral fraction. The contour 
levels and colour map have been chosen to closely mimic Figure~6 in both 
\cite{pawlikSchaye08} and \cite{pawlikSchaye11}. We have done this to 
highlight a key benefit that ray tracing codes such as \acro{} have over 
photon packet propagation methods such as TRAPHIC: isotropy. At the same 
$N=64^3$ particle resolution \acro{} is more spherically symmetric than 
TRAPHIC, even with their use of Monte-Carlo re-sampling. Furthermore, \acro{} 
outperforms TRAPHIC in this aspect even at early times (top panels in 
Figure~\ref{fig:stromslice}). Here the interior of the sphere is represented 
by 3.3 times fewer particles than the late time \strom{} spheres plotted in 
the TRAPHIC papers.

\begin{figure*}
\includegraphics[width=0.95\linewidth]{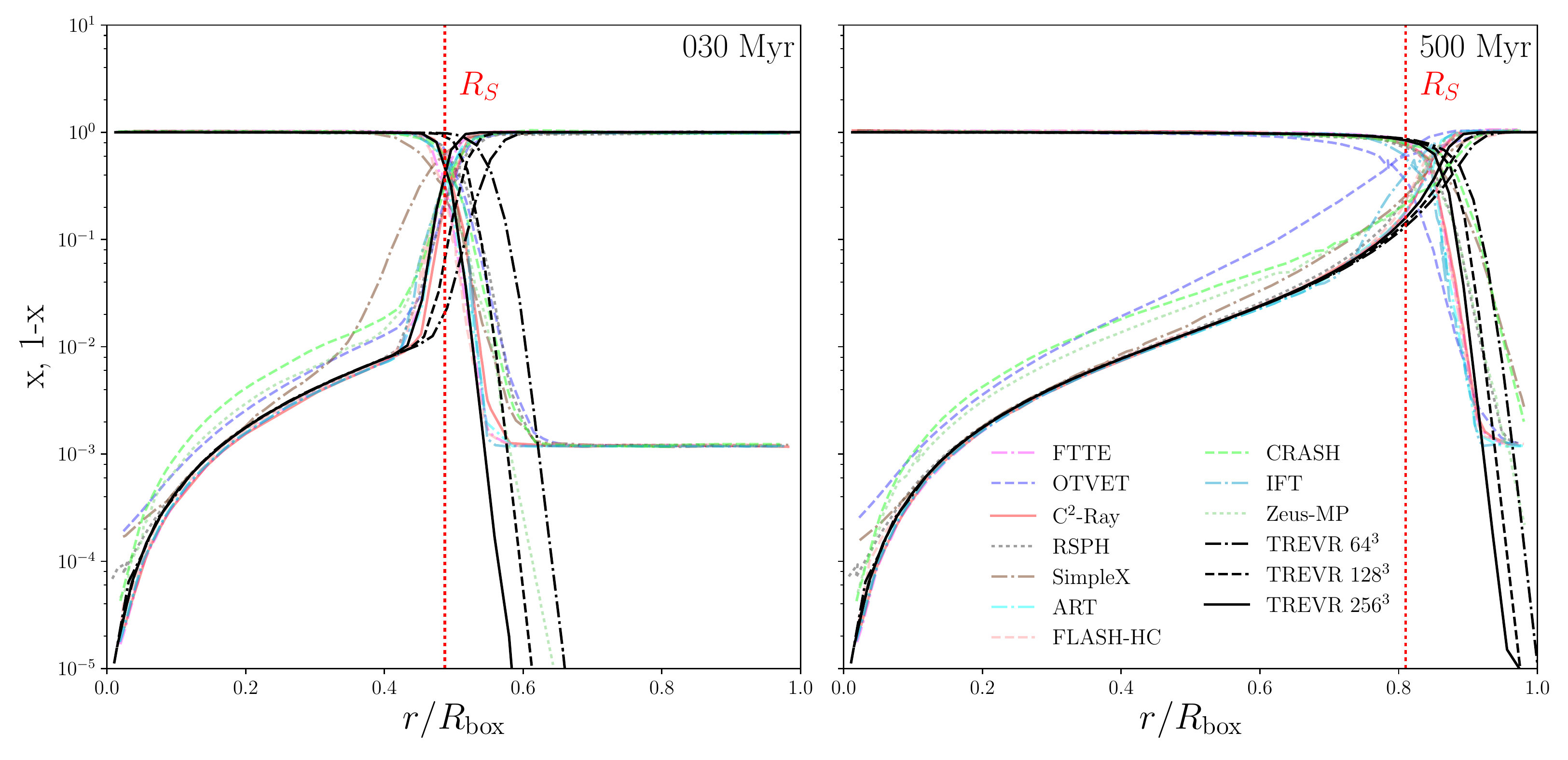}
\caption{Spherically averaged neutral and ionized fraction 
($\rm x$ and $\rm 1-x$) profiles for the isothermal \strom{} sphere test 
during the fast expansion (left) and equilibrium (right) stages. Radius on the 
$x$-axis is normalised by a box length of 6.6 kpc for comparison with plotted 
solutions from the \protect\cite{ilievEt06} comparison paper.}
\label{fig:stromiso}
\end{figure*}
Figure~\ref{fig:stromiso} is a plot of neutral/ionization fraction as a 
function of radius from the \strom{} sphere centre. The sharp \strom{} radius 
is plotted as well as non-sharp solutions from all codes presented in Figure~8 
of \cite{ilievEt06}. \acro{} tends to over-ionize at lower resolutions, but 
recreates the ionization profile quite well overall. At 30 Myr we converge 
with resolution to the sharp solution. At 500 Myr we converge to the non-sharp 
numerical solutions, which also over-ionize relative to the sharp solution at 
late times. Overall, the two higher resolution solutions are within the 
scatter of the non-sharp solutions of the codes presented in \cite{ilievEt06}. 

\subsubsection{The Non-Isothermal \strom{} Sphere}
The above test assumed the hydrogen gas was isothermal and that all incident 
photons had the same energy. In reality, photons range across many wavelengths 
with differing cross-sections at each wavelength. Absorption results in 
heating as well, which affects many gas properties including the recombination rate.

We reran the \strom{} sphere test, but this time the incident photons are 
assumed to be from a black body with a temperature of $10^5$ K. The 
cross-section is now photon energy weighted, giving 
$\sigma = 1.63 \times 10^{-18}$ $\rm cm^{-2}$. The gas has an initial 
temperature of 100 K and the recombination rate is a function of temperature 
set by
\begin{equation}
\alpha(T) = 2.59 \times 10^{-13} \left( \frac{T}{10^4 \hspace{5pt} 
{\rm K}}\right)^{-0.7} \hspace{5pt} {\rm cm}^{-3} \hspace{5pt} {\rm s}^{-1}
\end{equation}
to match \cite{petkovaSpringel09}. This test includes heating due to 
absorption and cooling due to recombination $\Delta_r$, collisional 
ionization $\Delta_{ci}$, line cooling $\delta_l$, and Bremsstrahlung 
radiation $\Delta_B$. The rates are taken from \cite{cen92} in order to
match \cite{petkovaSpringel09}.

\begin{figure*}
\includegraphics[width=0.95\linewidth]{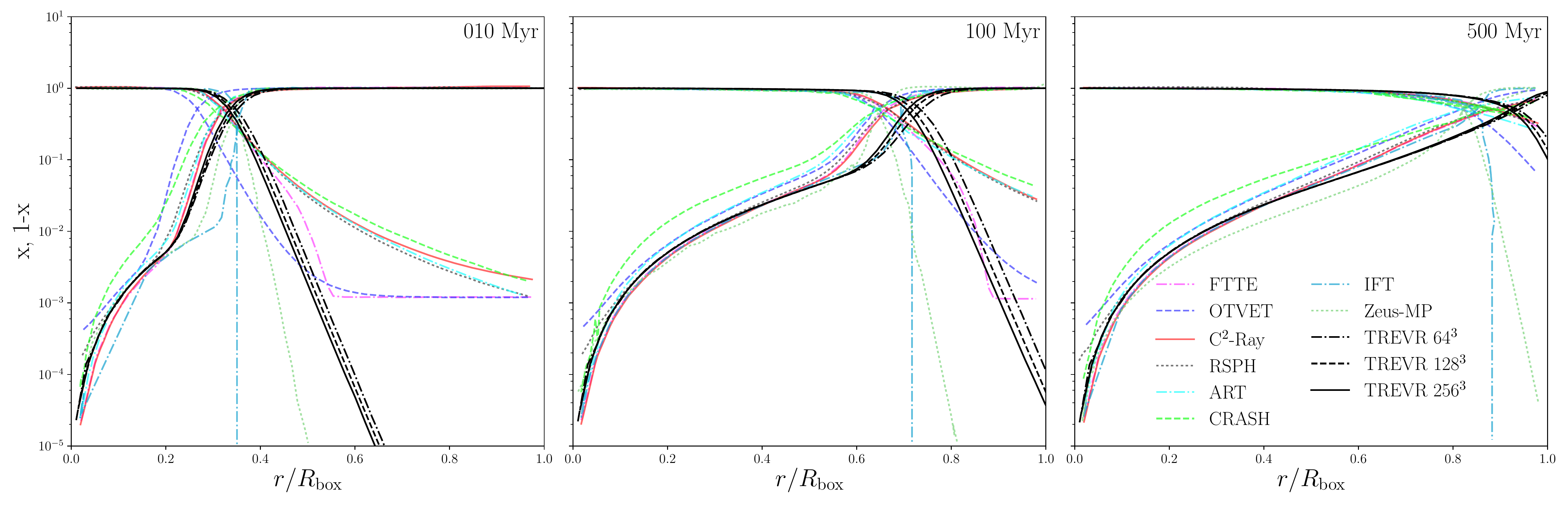}
\caption{Spherically averaged neutral and ionized fraction 
($\rm x$ and $\rm 1-x$) profiles for the non-isothermal \strom{} sphere test 
during the fast expansion (left), slowing down (middle) and equilibrium 
(right) stages.}
\label{fig:stromtherm}
\end{figure*}
\begin{figure*}
\includegraphics[width=0.95\linewidth]{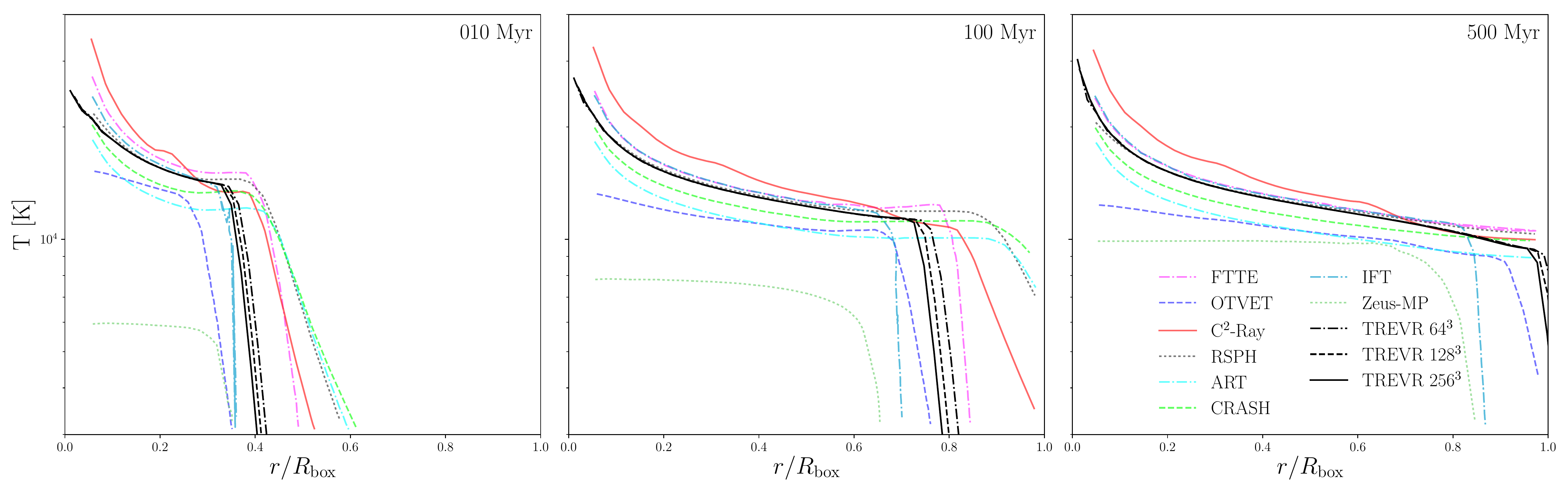}
\caption{Spherically averaged temperature profiles for the non-isothermal 
\strom{} sphere test.}
\label{fig:stromtemp}
\end{figure*}
Figures~\ref{fig:stromtherm} and~\ref{fig:stromtemp} show the neutral/ionized 
fraction and temperature respectively as a function of radius at $t=$ 10, 100
and 500 Myr. These times represent the fast expansion stage, slowing down 
stage and final equilibrium \strom{} sphere respectively. We have plotted 
numerical solutions from Figures~16 and~17 in \cite{ilievEt06} for comparison. 
Again, \acro{} recreates these profiles quite well.
\acro{} gives a somewhat large sphere radius which is 
due in part to the ionization code rather than the radiation method.  
The temperature profile lies in the middle of the scatter of the 
\cite{ilievEt06} solutions.  We note that for this test different codes employed
different assumptions about radiation bands and ionization treatments which
makes detailed comparisons difficult.

\section{Discussion And Conclusions}\label{sec:disc}
In this paper we have presented \acro{}, a practical and efficient, general 
purpose algorithm for computing RT in astrophysics simulations. For a RT 
method to be these things it must remain efficient with large numbers of 
resolution elements and radiation sources, compute the radiation field to a 
desired level of accuracy and handle density and opacity distributions 
representing the optically thick, thin and intermediate regimes.

\acro{}'s ability to scale feasibly with $N$ and $\NS$ is achieved by reducing 
all three of the cost multipliers of an $\bigO{\NK \NS N^{1/3}}$ naive ray 
trace:
\begin{enumerate}
\item Reverse ray tracing allows for the use of adaptive timesteps. The 
initial dependence on $\NK \sim N$ resolution elements is reduced to just the 
active radiation sinks (i.e. gas). $\NK$ is effectively hundreds of times smaller than 
$N$ when averaged over a large number of substeps.  
\item Source merging based on an opening angle criterion reduces the linear 
dependence on $\NS$ to $\log\NS$.
\item By adaptively reducing the resolution of rays via \acro{}'s novel 
refinement criterion, the $N^{1/3}$ cost of computing the optical depth along a 
ray can be reduced to $\log N$, while maintaining a specified level of accuracy.  
\end{enumerate}

In Section~\ref{sec:algo} we theoretically predicted \acro{}'s 
$\bigO{\NK \log\NS \log N}$ scaling behaviour. In the general case, 
represented by the perturbed glass test case with accuracy parameters of 
$\tr=0.1$ and $\tO=0.75$ (Figures~\ref{fig:openangle}, \ref{fig:refcrit} 
and~\ref{fig:pscale}), we have shown that \acro{} can indeed scale as 
predicted whilst achieving $\sim 1\%$ error. We also note that better than 
$\bigO{\NK\log\NS\log N}$ scaling (i.e. closer to $\bigO{\NK \log\NS}$) could 
be achieved for a medium with low optical depths via a more aggressive, 
top-down ray walk with our refinement criterion. 

The only general ray-tracing code we are aware of with similar scaling is 
TREERAY \citep{Wunsch2018}. TREERAY does not use an adaptivity criterion and has 
a fixed number of rays. This rather rigid approach has a benefit which is that 
the source and absorption walks can be combined to given an overall 
$\bigO{N log N}$ scaling, albeit without error controls and limited 
directional accuracy (e.g. for shadowing). TREERAY, as currently implemented 
in FLASH, uses a global timestep and is thus unable to take advantage of the 
large speed-ups reverse ray tracing can achieve via adapative timestepping.
However, this is not a limitation intrinsic to the TREERAY method itself.   

We note that the opening angle criterion is guaranteed to limit
errors in the low optical depth regime.  However, with absorption, it
sets an effective angular resolution below which shadows from distinct
sources would merge.  As shown in the tests (see particularly
sections~\ref{sinusoidtest}), this does not adversely effect the RMS
errors with absorption in cases with many sources.  However, it can adjust the
location of shadow edges (which is where most of the error resides).  
In the case of a few strong sources, the
user could employ a stricter merging criterion (or one based on
brightness) and have arbitrarily good angular resolution with
relatively little cost as demonstrated in section~\ref{spheretest}.

In plots of accuracy as a function of $\tr$ (Figures~\ref{fig:refcrit} 
and~\ref{fig:isosph}) we have also shown that \acro{}'s refinement criterion 
provides a predictable bound on accuracy, as we found the 
RMS relative error is $\propto \tr$ and the RMS 
errors do not exceed $\tr$.  

This behaviour enables \acro{} to reap the benefits inherent in instantaneous 
ray tracing methods whilst still being practical and general.  For example,
we can use any convenient timestep rather than being limited by the speed of 
light. Directional accuracy is another one of these benefits as is apparent in 
the sharp shadows cast in the isothermal spheres test 
(Figure~\ref{fig:cellplot}). Low levels of noise and anisotropy are also 
benefits compared with evolutionary methods as is apparent in the \strom{} 
sphere test (Figure~\ref{fig:stromslice}).

In the version of \acro{} as currently implemented, there are still some 
problems not easily handled. First, in any completely optically thick medium 
where high accuracy is required our method will result in worst case scaling 
of $\bigO{\NK\,N^{1/3}}$ (characterized in Section~\ref{sec:adref}). At face 
value this limits \acro{} to solving only post-reionization cosmology or 
similar problems that are largely optically thin. However, in optically thick 
media most sources contribute nothing to the local radiation field. In such 
cases \acro{} could easily terminate ray traces that are found (e.g. early in 
the optical depth sum when $\tau$ exceeds a threshold) or predicted (based on 
information from prior timesteps) to contribute little to no intensity to the 
final radiation field. These types of optimizations could also improve the 
weak scaling case.

A second problem is periodicity. Our method of a sphere of background sources 
providing a constant central background flux is adequate for isolated objects, 
but in the context of large cosmological boxes, such as reionization 
calculations, periodic boundaries are required. In such contexts, light travel 
times and redshifting are also potentially important. Such factors could be 
included in principle and this is a potential direction for future work.

Finally, there is the important issue of \textit{complex sources}. 
Consider a group of sources that meet the opening criterion and are merged, 
but are also contained within a region that has clumpy, opaque structures. 
Depending on the location of the merged centre of luminosity relative to the 
opaque clumps, the amount of radiation that escaped the merged source cell 
could vary significantly from that computed by the current algorithm.
Such cases would require that the opening criterion take the effect of
nearby absorbers into account, potentially using the information regarding
the variance in $\alpha$ already recorded for each cells. Such extended opening
and refinement criterion are the subject of ongoing investigations.

In addition to the above, future work could also include implementing 
scattering.  The process of scattering can be recast as an absorption, 
followed by an immediate re-emission of photons. Thanks to the $\log(\NS)$ 
scaling with radiation sources, this process can be implemented by considering 
resolution elements (SPH gas particles in our case) as sources of radiation 
without changing the scaling of the method.

A consequence of assuming an infinite speed of light is that radiation sinks 
will not see light as it was when emitted in the past but as the source appears
at the current time.  This is easy to remedy, as we have both the age of the 
source as well as the distance travelled by the photons. We can then 
\textit{age} the radiation sources with respect to the receiving resolution 
element, such that the received photons are representative of the luminous 
source as it was.

Currently, \acro{} only computes the radiation field in specific bands. 
\acro{} can handle many bands of radiation with a small constant multiplier 
added to the cost.  However, it may be advantageous to evolve the spectral 
shape over distance using an opacity which is a function of wavelength where the 
absorption is provided by a relatively simple or easily characterized set of 
species. This would also enable us to incorporate redshifting effects 
important for the evolution of large boxes over cosmological time periods.

J. Wadsley and H. M. P. Couchman would like to acknowledge the support of 
NSERC.


\bibliographystyle{mnras}
\bibliography{references}



\appendix
\section{Creating the Sinusoidally Perturbed Glass IC}
\label{sec:icnd}
To create our gently varying density distribution for the many source tests, 
we modify positions of particles in a glass IC by adding the 
sum of 24 sinusoidal modes to the initial particle positions as in 
Equation~\ref{eqn:manfft} below
\begin{equation}\label{eqn:manfft}
\vec{r} = \vec{r}_{\rm 0} + \sum_{i=1}^{24} \frac{1}{275} \sin 
\left( k_{x,i} r_x + k_{y,i} r_y + k_{z,i} r_z + \phi_i\right),
\end{equation}
where $\vec{r_0}$ is the particles initial position in the glass and $\vec{r}$ 
is its perturbed position in the final distribution. The $\vec{k_i}$ and 
$\phi_i$ values are listed in Table~\ref{tab:modes} in order to facilitate 
reproduction of the scaling tests. Both gas and star particles have the same 
density distribution. However, the initial glass was flipped for the star 
particles by reassigning $x$,$y$ and $z$ coordinates via
 \begin{equation}
 x_{\rm star} = y_{\rm gas}, \hspace{5pt} y_{\rm star} = z_{\rm gas}, 
 \hspace{5pt} z_{\rm star} = x_{\rm gas},
 \end{equation}
to prevent the particles from occupying the same position in space.

\begin{table}
	\centering
	\caption{Randomly generated $\vec{k}$ and $\phi$ values used in generating 
             the sinusoidally perturbed glass IC.}
	\label{tab:modes}
	\begin{tabular}{llllllll} 
		\hline
		$i$ & $k_{x,i}$ & $k_{y,i}$ & $k_{z,i}$ & $\phi_i$ \\
01 & $-$3.918398 & $+$1.727743 & $-$4.476095 & 0.829776 \\
02 & $-$3.681821 & $-$4.619688 & $+$4.865007 & 3.891157 \\
03 & $-$4.831801 & $+$3.769470 & $+$0.567451 & 3.668730 \\
04 & $-$2.298279 & $+$1.501757 & $+$4.716946 & 1.528348 \\
05 & $-$0.289974 & $-$3.097958 & $+$1.270028 & 4.113001 \\
06 & $+$1.262943 & $-$1.661726 & $-$2.600413 & 4.481799 \\
07 & $+$1.588224 & $+$4.072259 & $+$0.616444 & 2.971965 \\
08 & $-$2.253394 & $-$2.806478 & $+$2.749155 & 0.442241 \\
09 & $-$1.432569 & $+$3.324710 & $+$4.842991 & 2.871989 \\
10 & $+$1.287742 & $-$4.575517 & $-$4.001723 & 1.727810 \\
11 & $+$4.769704 & $+$0.540096 & $-$4.203839 & 5.872117 \\
12 & $-$3.013200 & $-$1.871251 & $-$2.514416 & 1.574008 \\
13 & $-$4.588620 & $+$4.384224 & $+$1.246849 & 1.985715 \\
14 & $-$0.372817 & $+$0.195243 & $+$4.074056 & 6.248739 \\
15 & $-$1.842232 & $+$0.901598 & $-$4.453613 & 6.273336 \\
16 & $+$1.986937 & $-$1.037650 & $+$1.958888 & 2.177783 \\
17 & $-$1.748485 & $-$1.386029 & $+$3.755833 & 0.532604 \\
18 & $+$4.852406 & $-$3.272506 & $+$0.826504 & 5.525470 \\
19 & $+$3.663293 & $-$4.597598 & $-$0.890135 & 4.528870 \\
20 & $-$1.720903 & $+$2.726011 & $+$3.192427 & 3.875610 \\
21 & $+$4.973332 & $+$4.777182 & $-$2.515792 & 0.406737 \\
22 & $+$0.057238 & $-$2.972427 & $-$1.828550 & 4.125258 \\
23 & $+$0.938234 & $-$0.487023 & $-$2.755097 & 1.335299 \\
24 & $+$1.943361 & $+$0.388178 & $-$3.783953 & 4.774938 \\
		\hline
	\end{tabular}
\end{table}

\bsp
\label{lastpage}
\end{document}